\documentclass[numberedappendix, appendixfloats]{emulateapj}
\usepackage{longtable, hyperref, graphicx, subfigure}
\usepackage[fleqn]{amsmath}
\usepackage{lmodern}
\usepackage[T1]{fontenc}
\usepackage{url}
\newcounter{column_number}
\newcommand{\Msun}{\ifmmode {M_{\odot}}\else${M_{\odot}}$\fi}
\newcommand{\Rsun}{\ifmmode {R_{\odot}}\else${R_{\odot}}$\fi}
\newcommand{\lapprox }{{\lower0.8ex\hbox{$\buildrel <\over\sim$}}}
\newcommand{\gapprox }{{\lower0.8ex\hbox{$\buildrel >\over\sim$}}}

\def\asec{\ifmmode^{\prime\prime}\else$^{\prime\prime}$\fi}

\def\ergs{$\mathrm{erg}$~$\mathrm{s}^{-1}$}
\def\ergcms{$\mathrm{erg}$~$\mathrm{cm}^{-2}$~$\mathrm{s}^{-1}$}
\def\ROSAT{\it ROSAT}
\def\CHANDRA{\textit{Chandra}}
\def\LL{$L_{\mathrm{X}}/L_{\mathrm{bol}}$}
\def\Prot{$P_{\mathrm{rot}}$}
\def\Pmem{$P_{\mathrm{mem}}$}

\slugcomment{DRAFT \today}
\shorttitle{Activity and Rotation in M37}
\shortauthors{N\'u\~nez et al.}

\begin{document}

\title{Linking Stellar Coronal Activity and Rotation at 500 Myr: A Deep Chandra Observation of M37}

\author{Alejandro N\'u\~nez\altaffilmark{1}, 
Marcel~A.~Ag{\"u}eros\altaffilmark{1}, 
Kevin~R.~Covey\altaffilmark{2},
Joel~D.~Hartman\altaffilmark{3},
Adam~L.~Kraus\altaffilmark{4},
Emily~C.~Bowsher\altaffilmark{1},
Stephanie~T.~Douglas\altaffilmark{1},
Mercedes~L\'opez-Morales\altaffilmark{5},
David~A.~Pooley\altaffilmark{6,7},
Bettina~Posselt\altaffilmark{8},
Steven~H.~Saar\altaffilmark{5},
Andrew~A.~West\altaffilmark{9}
}

\altaffiltext{1}{Department of Astronomy, Columbia University, 550 West 120th Street, New York, NY 10027, USA} 
\altaffiltext{2}{Department of Physics and Astronomy, Western Washington University, 516 High Street, Bellingham, WA 98225, USA}
\altaffiltext{3}{Department of Astrophysical Sciences, Princeton University, 4 Ivy Lane, Princeton, NJ 08544, USA}
\altaffiltext{4}{Department of Astronomy, University of Texas at Austin, 2515 Speedway, Stop C1400, Austin, TX 78712, USA}
\altaffiltext{5}{Harvard-Smithsonian Center for Astrophysics, 60 Garden Street, Cambridge, MA 02183, USA}
\altaffiltext{6}{Department of Physics, Sam Houston State University, Farrington Building, Suite 204, Hunstville, TX 77341, USA}
\altaffiltext{7}{Eureka Scientific, Inc., 2452 Delmer Street, Suite 100, Oakland, CA 94602, USA}
\altaffiltext{8}{Department of Astronomy and Astrophysics, Penn State University, 525 Davey Lab, University Park, PA 16802, USA}
\altaffiltext{9}{Department of Astronomy, Boston University, CAS 422A, 725 Commonwealth Ave, Boston, MA 02215, USA}

\begin{abstract}  
Empirical calibrations of the stellar age-rotation-activity relation (ARAR) rely on observations of the co-eval populations of stars in open clusters. We used the \CHANDRA\ {\it X-ray  Observatory} to study M37, a 500-Myr-old open cluster that has been extensively surveyed for rotation periods (\Prot). M37 was observed almost continuously for five days, for a total of 440.5 ksec, to measure stellar X-ray luminosities ($L_{\mathrm{X}}$), a proxy for coronal activity, across a wide range of masses. The cluster's membership catalog was revisited to calculate updated membership probabilities from photometric data and each star's distance to the cluster center. The result is a comprehensive sample of 1699 M37 members: 426 with \Prot, 278 with X-ray detections, and 76 with both. We calculate Rossby numbers, $R_o =\ P_{\rm rot}/\tau$, where $\tau$ is the convective turnover time, and ratios of the X-ray-to-bolometric luminosity, $L_{\mathrm{X}}/L_{\rm bol}$, to minimize mass dependencies in our characterization of the rotation-coronal activity relation at 500 Myr. We find that fast rotators, for which $R_o<0.09\pm0.01$, show saturated levels of activity, with log(\LL)$=-3.06\pm0.04$. For $R_o\geq0.09\pm0.01$, activity is unsaturated and follows a power law of the form $R_o^{\beta}$, where $\beta$=$-2.03_{-0.14}^{+0.17}$. This is the largest sample available for analyzing the dependence of coronal emission on rotation for a single-aged population, covering stellar masses in the range 0.4--1.3 \Msun, \Prot\ in the range 0.4--12.8 d, and $L_{\mathrm{X}}$ in the range 10$^{28.4-30.5}$ \ergs. Our results make M37 a new benchmark open cluster for calibrating the ARAR at ages of $\approx$500 Myr.
\end{abstract}


\section{Introduction}
X-ray emission traces the strength of a stellar magnetic dynamo and is tightly linked to stellar rotation and age. Rotational shear at the boundary between the radiative and convective zones is thought to power the solar dynamo \citep[e.g.,][]{parker93}. The generated magnetic fields heat the corona to temperatures sufficient to produce X rays \citep{parker79, cowling81}. Observations of solar-like main-sequence (MS) stars confirm this connection between rotation and X-ray emission, finding that the X-ray luminosity $L_{\mathrm{X}} \propto v_{\mathrm{rot}}$$^\beta$, where $v_{\mathrm{rot}}$ is the stellar surface rotational velocity and $\beta\approx2$ \citep{palla81, pizzo03}. 

In addition, observations have shown that stars rotate more slowly as they age, with rotation decaying with time such that $v_{\mathrm{rot}} \propto t^{-\alpha}$, with $\alpha=0.5$ for solar-mass stars \citep[e.g.,][]{skumanich72,meibom2011}. Angular momentum lost through stellar winds is generally thought to be responsible for the Skumanich law, with the exact dependence of $v_{\mathrm{rot}}$ on age being a function of the assumed field geometry and degree of core-envelope coupling \citep{kawaler1988, krish1997}. 

Combined, these two effects predict that for solar-type stars, $L_{\mathrm{X}}$ should decay such that $L_{\mathrm{X}} \propto t^{-\alpha\beta}$, with $\alpha\beta=1$. It is also known, however, that this picture of a uniformly decaying $L_{\mathrm{X}}$ does not apply to stars with ages $\lapprox$100 Myr. Surveys of solar-type stars in the Orion Nebula and the Pleiades ($\lapprox$10 and 100 Myr, respectively) found that $L_{\mathrm{X}}$ decays more slowly: $L_{\mathrm{X}} \propto t^{-0.75}$ \citep{queloz98, preibisch2005}. Because of the paucity of older, nearby clusters, constraints are more limited at $\gapprox$200 Myr and come mostly from field stars \citep[e.g.,][]{gudel2007}: observations of five solar analogs indicated that $L_{\mathrm{X}} \propto t^{-1.5}$ for $t$ > 1 Gyr \citep{gudel1997}, as did a survey of the 1--2 Gyr cluster NGC 752 \citep{giardino2008}. A change in the magnetic field topology or core-envelope decoupling are the commonly invoked explanations for this sharp drop-off in $L_{\mathrm{X}}$, but it remains poorly understood. Chromospheric activity, also thought to trace the strength of the stellar dynamo, may also suffer a similarly steep decline at ages $>$1 Gyr \citep{pace2004}. 

The homogeneous, co-eval populations of open clusters provide an ideal environment for calibrating empirically the stellar age-rotation-activity relationship (ARAR). The Hyades's proximity ($\approx$50 pc) has facilitated observations of its members at many wavelengths and established it as the benchmark middle-aged (500-600 Myr) cluster \citep{perryman1998}. Empirical calibrations of the ARAR often interpolate directly between the observed properties of solar-type Hyads and that of the Sun. The Hyades therefore serves as the only anchor for our understanding of coronal emission between 500-600 Myr and 4.5 Gyr, a significant fraction of a solar-type star's MS lifetime.

However, the Hyades's proximity is a drawback for X-ray studies. The cluster spans  well over 10$^{\circ}$, and modern X-ray telescopes with small fields of view are poorly suited to large-scale studies of the cluster. Indeed, the most recent such study was conducted by \citet{stern1995}, who detected 185 Hyads with {\it ROSAT}.

Recently, \citet{Douglas14} probed the rotation-activity relation for stars in the Hyades and Praesepe, another $\approx$600-Myr-old cluster, using both coronal and chromospheric ($L_{\mathrm{H}\alpha}$) activity measurements. These authors found that coronal activity declines faster than chromospheric activity as a function of (unsaturated) Rossby number $R_o =\ P_{\rm rot}/\tau$, where $\tau$ is the convective turnover time. However, because of the paucity of stars in either cluster with measured rotation periods (\Prot) and X-ray detection (24 in the Hyades and 15 in Praesepe), the exact relationship between coronal activity and rotation at this age remains elusive. 

Other studies \citep[e.g.,][]{palla81, gudel1997, Randich2000b, mamajek2008, wright2011} have examined the stellar rotation-coronal activity relation, and generally suffer from similar issues: either there are too few stars at a given age to get significant statistics or the population is too heterogeneous to strongly constrain age-dependent properties.

M37 (NGC 2099), a Hyades analog at $\approx$1.5 kpc, has been extensively surveyed in the optical by \citet{messina08a} (hereafter ME08), \citet{hartman08a} (hereafter HA08), and \citet{hartman09} (hereafter HA09), resulting in hundreds of \Prot\ measurements. As part of our Palomar Transient Factory Open Cluster Survey \citep[][]{agueros11}, we obtained a complementary deep (440.5 ksec) {\it Chandra} observation of the cluster. M37 is distant enough that a significant number of cluster stars fall within the footprint of {\it Chandra}'s ACIS-I chips. Our X-ray data, combined with M37's rich membership and the extensive set of available \Prot\ data, make it an excellent laboratory for examining the dependence of coronal activity on rotation. 

We begin in Section~\ref{observations} by describing our \CHANDRA\ data and data reduction, and the construction of our X-ray catalog. In Section~\ref{optical}, we consolidate the ME08 and HA09 \Prot\ data for the cluster, and we re-visit the cluster membership using our own membership probability calculations (fully described in Appendix~\ref{membershipcalc}.) In Section~\ref{analysis}, we derive  stellar quantities using optical and X-ray data, and we match our X-ray sources to the consolidated list of optical objects. In Section~\ref{results}, we present our results and compare them to those in the literature. We conclude in Section~\ref{conclusion}. We present spectral and timing analysis of our highest-count X-ray \CHANDRA\ sources in Appendix~\ref{spectra}.

\section{{\it Chandra} Observations}\label{observations}
\begin{figure}
\includegraphics{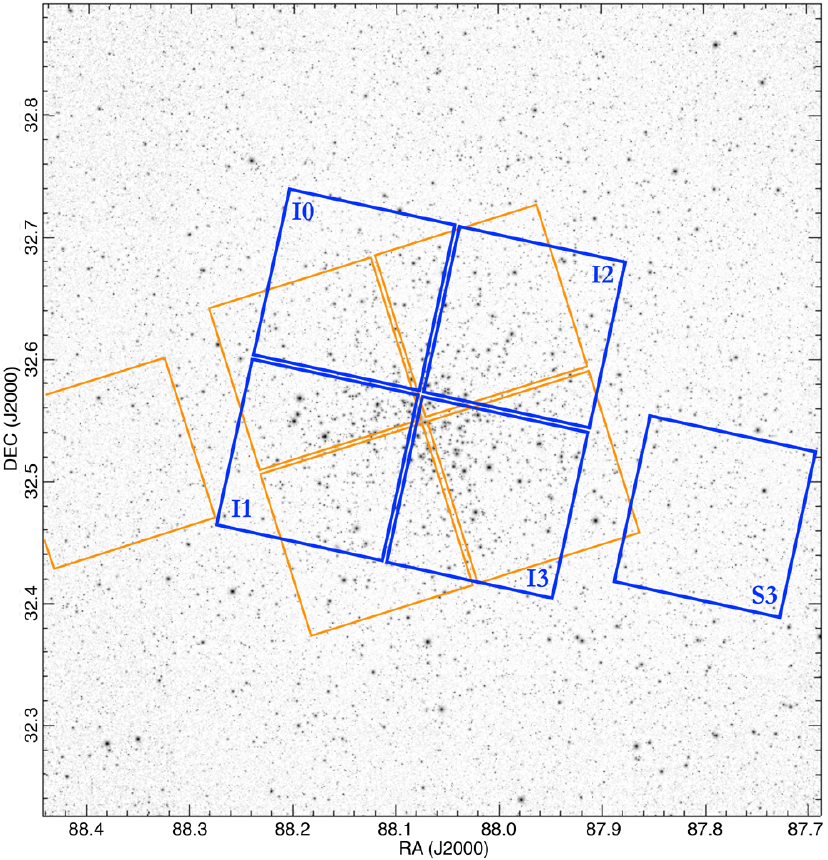}
\caption{Thick blue squares (with chip numbers) indicate the footprint of the \CHANDRA\ ACIS chips for four of our observations; the thin orange squares indicate that of our fifth observation. A 40$\arcmin$$\times$40$\arcmin$ $i'$ Megacam image of M37 from \citet{hartman08a} serves as the background.}
\label{fig:optical}
\end{figure}

The central field of M37 was observed five separate times between 2011 Nov 14 20:58 and 2011 Nov 19 15:31 UTC for a total of 440.5 ksec with the Advanced CCD Imaging Spectrometer \citep[ACIS;][]{garmire03}. The four ACIS-I chips and the ACIS-S3 chip were used in Very Faint telemetry mode to improve the screening of background events and thus increase the sensitivity of ACIS to faint sources \citep{vikhlinin01}. The exposure-weighted average aimpoint of the 16$\farcm$9$\times$16$\farcm$9 ACIS-I field-of-view is $\alpha=05^{\mathrm{h}}52^{\mathrm{m}}17.^{\mathrm{s}}86$, $\delta=+32^{\circ}33'48\farcs23$ (J2000). The pitch angle for four observations was 103$^{\circ}$;  due to scheduling constraints, it was 253$^{\circ}$ for the fifth. Table \ref{tbl:obslog} provides the basic information for our five \CHANDRA\ observations and Figure \ref{fig:optical} shows their footprints superimposed on a 40$\arcmin$$\times$40$\arcmin$ $i'$ image centered on M37 obtained by HA08 with the Megacam on the MMT telescope.

\begin{deluxetable}{rcrccr}
\centering 
\tabletypesize{} \tablewidth{0pt}

\tablecaption{ Log of {\em Chandra} Observations
 \label{tbl:obslog}}

\tablehead{
\colhead{ObsID} & 
\colhead{MJD} & 
\colhead{Exposure\tablenotemark{a}} & 
\multicolumn{2}{c}{Nominal Aimpoint\tablenotemark{b}} & 
\colhead{Roll}  \\
\cline{4-5}
\colhead{} & 
\colhead{} & 
\colhead{(s)} & 
\colhead{$\alpha_{\rm J2000}$} & 
\colhead{$\delta_{\rm J2000}$} & 
\multicolumn{1}{c}{ (\arcdeg)}  
\setcounter{column_number}{1}
}

\startdata
13655 & 55879 & 51179 & 05:52:17.61 & +32:33:59.0 &  103 \\
13656 & 55886 & 165493 & 05:52:17.61 & +32:33:59.0 &  103 \\
13657 & 55941 & 92172 & 05:52:18.82 & +32:33:05.0 &  253 \\
14366 & 55882 & 65074 & 05:52:17.61 & +32:33:59.0 &  103 \\
14367 & 55883 & 60813 & 05:52:17.61 & +32:33:59.0 &  103 
\enddata

\tablenotetext{a}{Exposure times are the net usable times after various filtering steps are applied in the data reduction process.}

\tablenotetext{b}{The time-averaged location of the optical axis. The aimpoints are obtained from the satellite aspect solution before astrometric correction is applied. Units of right ascension are hours, minutes, and seconds; units of declination are degrees, arcminutes, and arcseconds.}

\end{deluxetable}

The data reduction was done using the \CHANDRA\ Interactive Analysis of Observations \citep[CIAO;][we used CIAO 4.6 and CALDB 4.6.1.1]{fruscione2006} tools. We began by using \texttt{chandra\_repro}, which executes all recommended data processing steps on the level 1 data, including subpixel randomization using the EDSER algorithm. This tool creates a new bad pixel file by identifying known bad pixels and pixels with bad bias values, searching for afterglows and hot pixels, and marking pixels adjacent to afterglows and hot pixels. \texttt{chandra\_repro} also creates a new level 2 event file that includes only events with grades 0, 2, 3, 4, 6, and a status of 0. To avoid excluding potentially good events in modestly bright point sources, we did not use the very-faint correction in \texttt{chandra\_repro}.

\subsection{X-ray Source Extraction}\label{extraction}
Our source extraction procedure was divided into two steps. First, we merged our observations using CIAO and ran \texttt{wavdetect} \citep{freeman02} with a liberal significance threshold to obtain a raw candidate source list. Second, we used the ACIS Extract (AE) point-source analysis software \citep[][we used AE Version 2014feb17]{broos10}\footnote{Available from \url{http://www2.astro.psu.edu/xray/docs/TARA/ae_users_guide.html}} to prune spurious detections from our candidate list and extract photometry. AE's algorithms are tailored to calculate source properties accurately when multiple observations are merged and analyzed.

\subsubsection{Merging the Observations}\label{merging}
To construct the merged event file, we first ran \texttt{wavdetect} on the observations using a false-positive probability threshold of 10$^{-6}$. We then used \texttt{reproject\_aspect} to register the observations to the astrometric frame of the longest exposure (ObsId 13656), using only source regions with \texttt{SRC\_SIGNIFICANCE} > 7, a 3$\arcsec$ matching radius, and a residual rejection limit (a parameter used to remove source pairs based on pair positional offsets) of 0$\farcs$6. The values of these parameters were chosen to have \texttt{reproject\_aspect} work with a more stringent source list \citep[see e.g.,][]{xue11}. The linear translations ranged from $-0\farcs38$ to $0\farcs63$, rotations ranged from $-0^{\circ}.006$ to $0^{\circ}.018$, and scale changes ranged from $0.9997$ to $1.0002$.

We used the CIAO tool \texttt{merge\_obs} to produce a merged event file, a combined exposure map, and exposure-corrected flux images. We constructed flux images and exposure maps for three bands: 0.5--7 keV (full), 0.5--2 keV (soft), and 2--7 keV (hard).

\begin{figure}
\begin{center}
\includegraphics[scale=0.195]{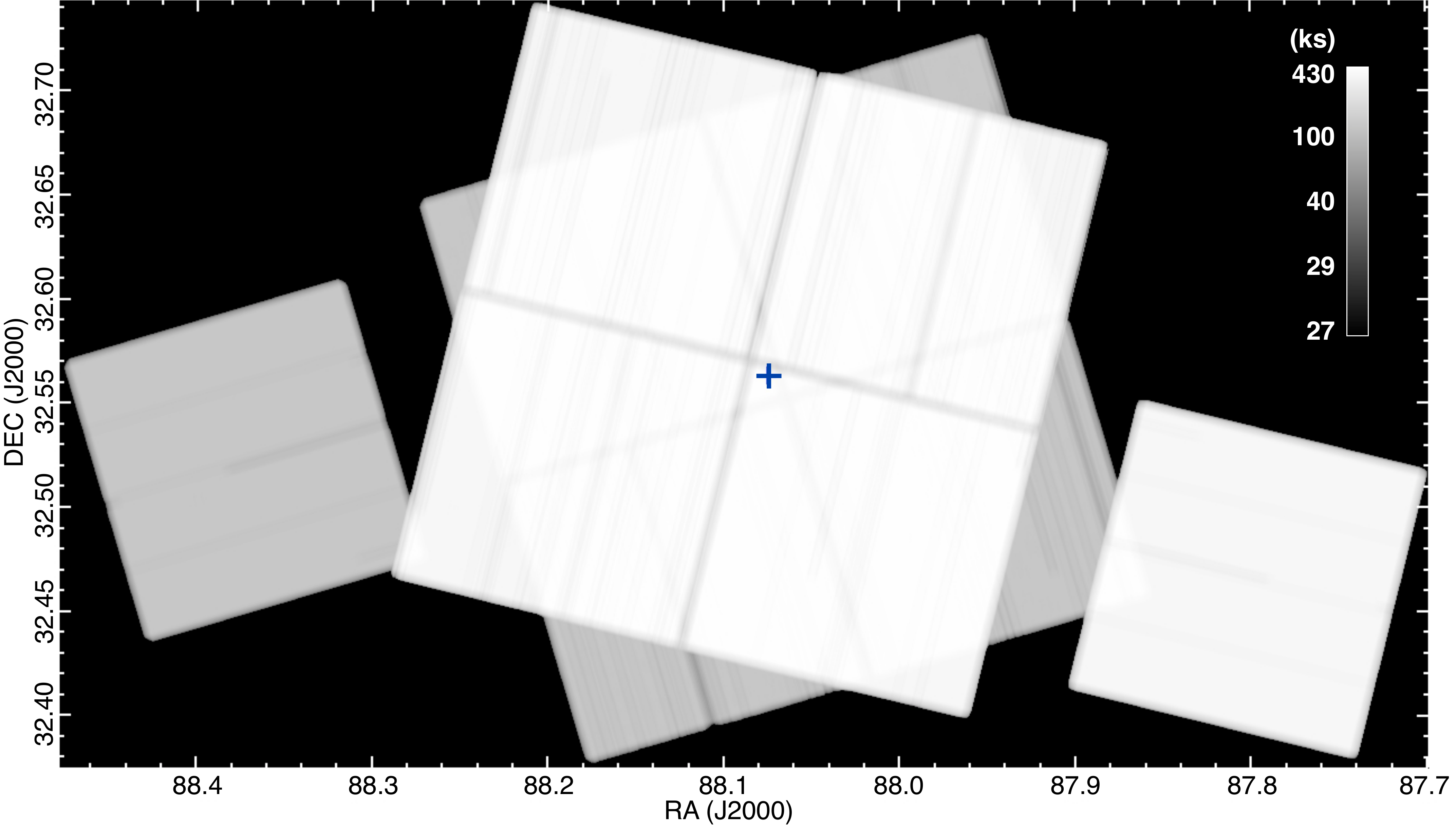}
\caption{Effective exposure map for the full (0.5--7 keV) band of our merged \CHANDRA\ event displayed in logarithmic gray scale (in units of ksec). The average aimpoint is indicated by the plus sign.}
\label{fig:expmap}
\end{center}
\end{figure}

\begin{figure*}
\begin{center}
\includegraphics[scale=0.4]{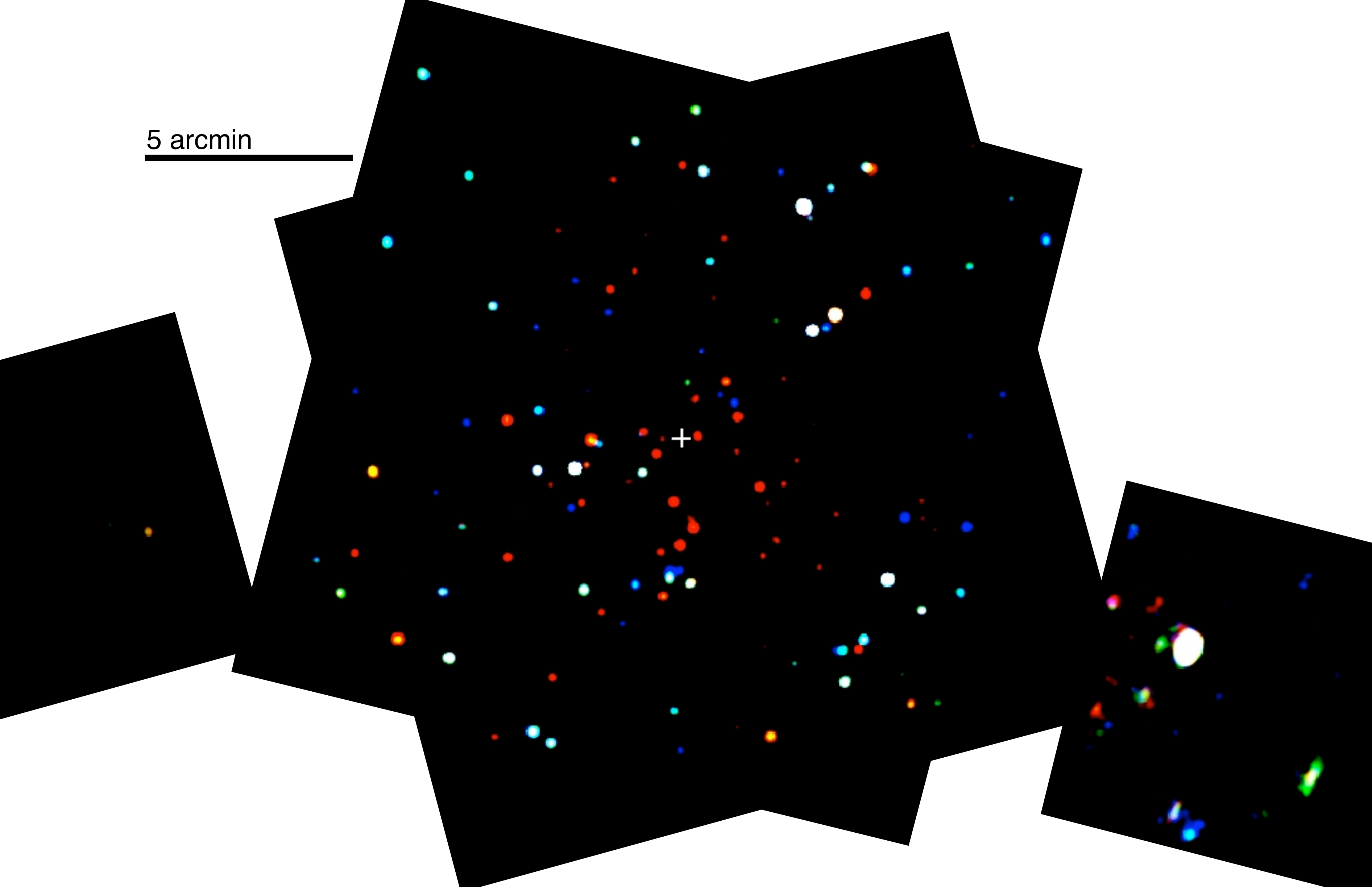}
\caption{\CHANDRA\ smoothed RGB composite image of the central region of M37. We combined, overbinned, and smoothed (using a Gaussian of kernel radius 5) the 0.5--1.5 keV (red), 1.5--2.5 keV (green), and 2.5--7 keV (blue) exposure-corrected flux images from our merged \CHANDRA\ event file. The plus sign indicates the average aimpoint.} 
\label{fig:colorcomposite}
\end{center}
\end{figure*}

Figure~\ref{fig:expmap} shows the full band effective-exposure map for the merged 440.5 ksec event. Since one of our five observations is rotated 150$^{\circ}$ relative to the others, the merged exposure map is inhomogeneous, with the longest exposure near and around the average aimpoint. We used \texttt{ds9} to produce a representative color composite image by combining, overbinning, and smoothing (using a Gaussian of kernel radius 5) the 0.5--1.5 keV (red), 1.5--2.5 keV (green), and 2.5--7 keV (blue) exposure-corrected flux images, and the result is shown in Figure~\ref{fig:colorcomposite}.

\subsubsection{Source Detection and Pruning}\label{detection}
To obtain an initial list of candidate X-ray sources, we first created point-spread function (PSF) maps for the individual observations using \texttt{mkpsfmap} with an encircled counts fraction (the fraction of the PSF to enclose) of 0.9. We then selected the minimum PSF map size out of the five observations at each pixel using \texttt{dmimgfilt}, which allowed us to detect point sources that are smaller than the mean size, but still larger than the local PSF in the individual maps. 

To obtain an initial candidate source list, we ran \texttt{wavdetect} using this minimum PSF map and the merged event file and exposure map. We used a scale sequence ranging from 1 to 8$\sqrt{2}$\ for the ACIS-I chips and from 8 to 32$\sqrt{2}$\ for ACIS-S3;\footnote{Larger scales are needed for  ACIS-S3 because of the large PSFs far from the aimpoint.} in both cases the scales had factor increments of $\sqrt{2}$. \citet{alexander01} pointed out that real sources can be inadvertently lost using a too stringent false-positive probability threshold. We therefore reset our threshold to $10^{-5}$ in \texttt{wavdetect} to allow for a non-negligible number of spurious sources to be included in the initial candidate list. 

Separate candidate source lists were prepared for the full, soft, and hard energy bands. To consolidate the three lists, we matched these lists against each other using a 2$\arcsec$ matching radius for sources with angular separation $\theta$ < 5$\arcmin$ from the average aimpoint, and 4$\arcsec$ for sources with $\theta$ > 5$\arcmin$. No multiple matches occurred using this approach. Once matched, the central location of each point source was adopted, in order of priority, from the full, soft, or hard energy band list. The resulting list consists of 794 candidate X-ray sources: 694 detected in the soft band (53 only in the soft band), 173 in the hard band (10 only in the hard band), and 608 in the full band (50 only in the full band).

To prune spurious sources from this list, we used AE's binomial probability \textit{P} \citep[][Appendix~A2]{weisskopf2007}.\footnote{\texttt{prob\_no\_source}, see section~5.10.3 of the AE manual, \url{http://www2.astro.psu.edu/xray/docs/TARA/ae_users_guide.pdf}.} \textit{P} is the probability that, based on the source and local background measurements, a candidate source is a background fluctuation. This approach to producing a final source catalog is preferable to implementing \texttt{wavdetect} with a more stringent false-positive probability threshold \citep[e.g.,][]{lehmer09, xue11, wright2014}. Not only is the pruning based on a transparent mathematical criterion, \textit{P}, but AE also uses a more complex approach to treat source regions in multiple images by replacing \texttt{wavdetect}'s ellipses with polygons that more closely simulate the PSF, and by calculating more accurate source positions, which maximizes the signal-to-noise ratio. We followed AE's suggestion and rejected candidate sources with \textit{P}~$\geq0.01$ as spurious. 

To refine the source positions, we improved those calculated by \texttt{wavdetect} using AE's source positions estimates. For source regions with $\theta\le5\arcmin$ we used the ``mean data'' position estimate, which is simply the mean position of events. For source regions with $\theta$ > 5$\arcmin$, we used the ``correlation'' position estimate, which determines the source center by correlating the neighborhood around the source (not just the extracted counts) with the PSF of the source. This is a more accurate source center position when the PSF is asymmetric, which is the case for source regions far off-axis.\footnote{See section~5.3 of the AE manual, \url{http://www2.astro.psu.edu/xray/docs/TARA/ae_users_guide.pdf}, for a discussion of source center estimates.}

The absolute astrometry of the individual observations was then corrected by using \texttt{reproject\_aspect} again to register them to the astrometric frame of the 2MASS Point Source Catalog \citep{cutri03}. This time, the procedure used the refined source positions of our X-ray sources and a 2MASS list of point sources of photometric quality A\footnote{Source confusion was not an issue for the 2MASS sources in the list we used.} in the {\it JHK}$_{\mathrm{s}}$ bands within a square of side 20$\arcmin$ centered on the ACIS-I average aimpoint ($\approx$1900 objects). We used a 2$\farcs$5 matching radius and a residual rejection limit of 0$\farcs$6. When multiple matches occurred, the closest one was selected as the true match. The linear translations ranged from $-0\farcs28$ to $-0\farcs16$, and a rotation angle of $-0^{\circ}.01$ and scale factor of $1.0003$ was applied to all five observations.

We iterated over the process of source and local background extraction, \textit{P} calculation, candidate source list purging of $P\geq0.01$ sources, source position refinement, and absolute astrometric correction, until no more sources were discarded. This was necessary because source and local background extraction in surviving sources may be affected by the disappearance of nearby purged sources. Our final \CHANDRA\ catalog contains 774 sources. Of these, 321 were detected in the three energy bands, four were detected in the full band only, and twelve in the soft band only. Almost half of the discarded sources were originally detected only in the full band. 

\begin{figure*}
\begin{center}
\includegraphics[scale=0.4]{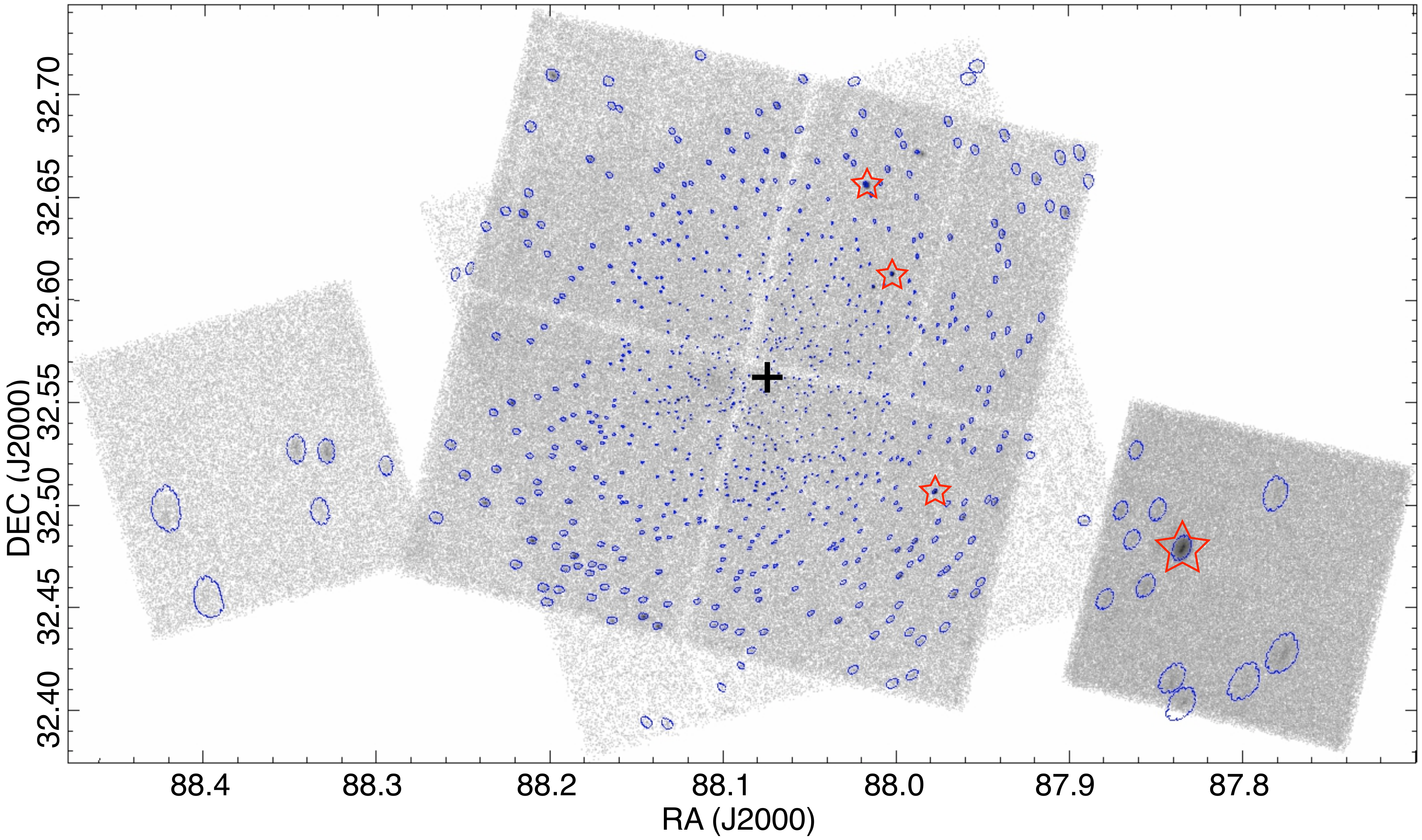}
\caption{Merged full-band (0.5--7 keV) flux image smoothed using a Gaussian function of kernel radius 4, with the polygons used for extracting source events, photometry, and spectra drawn in blue. The average aimpoint is indicated by the plus sign. Red stars indicate the four bright sources for which we perform spectral and timing analysis in Appendix \ref{spectra}. We identified 774 X-ray sources. The varying sizes of the source regions illustrate the degrading of the PSF far from the aimpoint.}
\label{fig:sourceregions}
\end{center}
\end{figure*}

\begin{figure}
\includegraphics{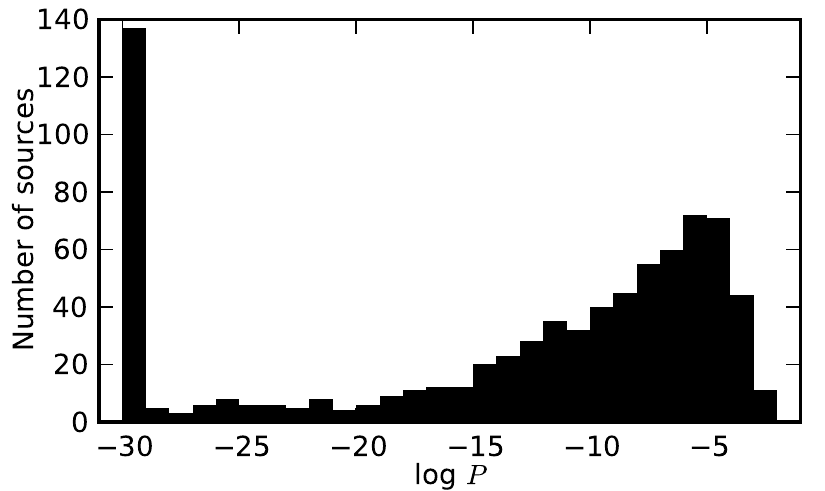}
\caption{AE's binomial probability $P$ that a source is just a background fluctuation for all sources that have $P$ < 0.01 in at least one of the three energy bands. We considered sources with $P\geq$ 0.01 to be spurious detections. For clarity, $P$ values lower than $10^{-30}$ are set to $10^{-30}$.}
\label{fig:PBs}
\end{figure}

The regions used to extract these sources are shown in Figure~\ref{fig:sourceregions} drawn on the merged full-band flux image, smoothed using a Gaussian function of kernel radius 4. Figure~\ref{fig:PBs} shows the distribution of the binomial probability for all the sources in our catalog. This histogram clearly shows that most sources have negligible probabilities of being false detections. The final merged event file has an absolute astrometry accuracy of $<$0$\farcs$63, based on 210 matches to 2MASS. We investigate the positional uncertainty of the individual X-ray sources in more detail in Section~\ref{uncertainty}.

\subsubsection{Extracting X-ray Photometry}\label{xphotometry}
We followed AE's recipe for photometry extraction\footnote{See \url{http://www2.astro.psu.edu/xray/docs/TARA/ae_users_guide/procedures/photometry_procedure.txt}} using the three energy bands (full, soft, and hard) described in Section~\ref{merging}.
AE first defines each extraction region with a polygon that approximates the $\approx$90\% encircled-energy fraction contour of the local PSF measured at 1.497 keV. It then calculates exposure times, source counts, and background counts for each region and energy band.\footnote{For background counts, we used AE's \texttt{BETTER\_BACKGROUNDS} algorithm, which models several background components simultaneously. See section~7.6.1 of the AE manual, \url{http://www2.astro.psu.edu/xray/docs/TARA/ae_users_guide.pdf}.}

AE executes these steps on each \CHANDRA\ observation independently and then merges the extraction data from each source by optimizing the source's signal-to-noise ratio. AE then computes photon fluxes using net (i.e., background-subtracted) source counts, exposure times, and ancillary response files generated by the CIAO tool \texttt{mkarf}.

\begin{figure}
\includegraphics{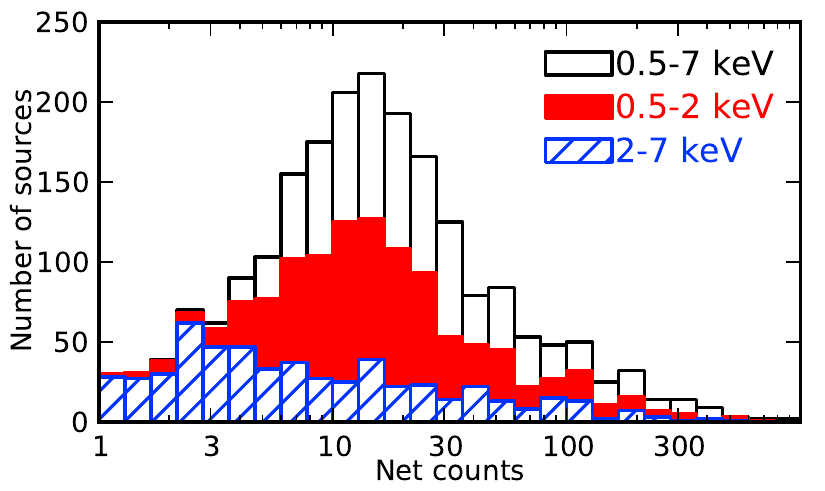}
\caption{Cumulative distribution of sources in our catalog based on their net counts in the full (0.5--7 keV), soft (0.5--2 keV), and hard (2--7 keV) bands.}
\label{fig:counts_hist}
\end{figure}

\begin{figure}
\includegraphics{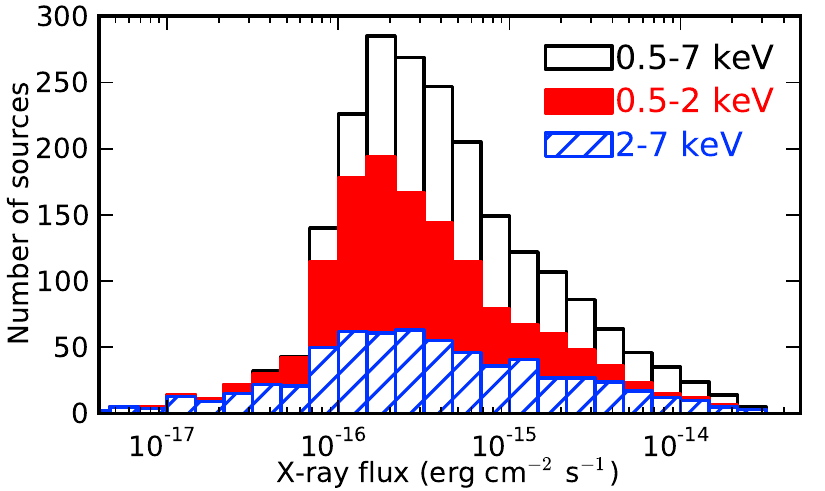}
\caption{Cumulative distribution of sources in our catalog based on their absorbed X-ray fluxes in the full (0.5--7 keV), soft (0.5--2 keV), and hard (2--7 keV) bands.}
\label{fig:fluxes_hist}
\end{figure}

Figures~\ref{fig:counts_hist} and \ref{fig:fluxes_hist} show the net counts and absorbed X-ray flux distributions in the three energy bands for all the sources in our catalog. We calculated the latter by multiplying the band photon flux by its median photon energy. The median number of net counts for the full, soft, and hard band is $\approx$19, 14, and 3, respectively. There are 78 sources with more than 100 net counts in the full band. In general, hard band detections are less common than full band detections, and most hard band detections have fewer than ten net counts. The median absorbed X-ray flux for the full, soft, and hard band is $4.8\times 10^{-16}$, $2.4\times 10^{-16}$, and $3.0\times 10^{-16}$ \ergcms, respectively.

\subsection{Estimating the Source Positional Uncertainties}\label{uncertainty}
To parametrize the dependence of the X-ray positional uncertainty on $\theta$ and source counts, we applied the approach of \citet{kim07}, who described the positional uncertainty $\Delta_x$ as
\begin{equation}\label{posunc}
    \textrm{log}\Delta_x = a_0 + a_1\theta + a_2\textrm{log}\ C,
\end{equation}
where $\Delta_x$ is in arcseconds, $\theta$ is the off-axis angle in arcminutes, $C$ is the net full-band counts, and $a_0$, $a_1$, and $a_2$ are constants. This parametrization takes into account the fact that the statistical uncertainty of the calculated PSF center of a \CHANDRA\ source increases with lower source counts, and that the size of the PSF increases with larger $\theta$. We performed our own parametrization of the relation above because the one by \citet{kim07} was based on \texttt{wavdetect}-derived positions and photometry.

We started by doing a preliminary match of our X-ray sources to objects in the optical survey of HA08, which is registered to the 2MASS astrometric frame, using 2$\farcs$5 and 4$\farcs$0 matching radii for X-ray sources with $\theta < 5\arcmin$ and > 5$\arcmin$, respectively. We selected these separations by inspecting a histogram of the number of matches as a function of angular separation, and selecting the values that resulted in a mismatch probability $<$1\%.  We excluded likely false matches based on full-band counts, $\theta$, and matching radius (e.g., a source with $>$100 net counts, $\theta<5$, and matching offset $>$1$\asec$). When multiple matches occurred, the closest one was selected as the true one. Using these criteria, we identified 452 X-ray sources as having optical counterparts. In Section~\ref{matching} we use a more rigorous matching technique that considers the derived positional uncertainty of each X-ray source from this section to create our final list of X-ray-to-optical matches.

Next, we performed a multivariate $\chi^2$ minimization of Equation~\ref{posunc} using the offsets between our X-ray sources and HA08 optical objects as proxy for positional uncertainties. Once we obtained values for $a_0$, $a_1$, and $a_2$, we adjusted the value of $a_0$ until $\gtrsim$80 \% of our X-ray sources with optical counterparts had $\Delta_x$ values larger than the match offsets. The final values are $a_0 = -0.2928$, $a_1=0.1200$, and $a_2=0.3401$.

\begin{figure}
\includegraphics{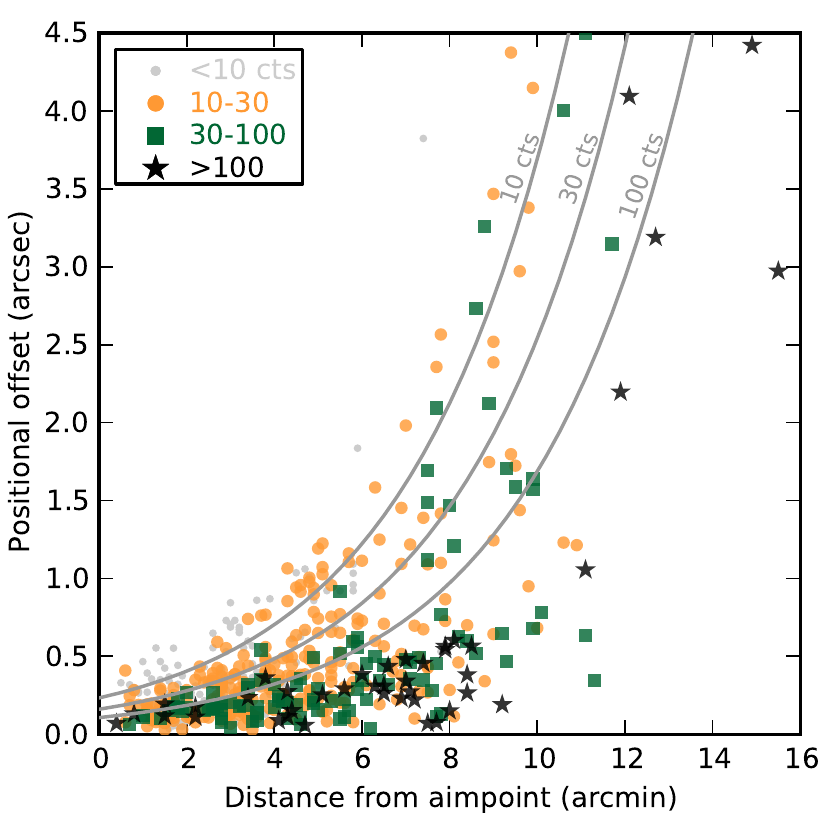}
\caption{Positional offset for our X-ray sources with optical counterparts in the HA08 survey, shown as a function of distance $\theta$ from the \CHANDRA\ aimpoint. Black stars are X-ray sources with >100 net counts in the full band; green squares, between 30 and 100, orange circles, between 10 and 30; and small gray circles, <10. We indicate $\approx$80\% positional uncertainty curves for sources with 10, 30, and 100 net counts following Equation~\ref{posunc}.}
\vspace{0.05 in}
\label{fig:theta_v_offset}
\end{figure}

We then used these values in Equation~\ref{posunc} to calculate the $\approx$80\% confidence positional uncertainties for all our sources, and we report these values in Table~\ref{tbl:X_sources}. Figure~\ref{fig:theta_v_offset} shows the positional offset as a function of $\theta$ for the X-ray-to-optical matches just described. We overplot three curves generated from Equation~\ref{posunc} using $C$ = 10, 30, and 100 counts.

To estimate the expected number of false matches in our catalog, we shifted the X-ray source positions by 30$\arcsec$ out to 6$\arcmin$ in all directions, and then re-matched them to the optical positions using a 2$\farcs$5 matching radius. The 2$\farcs$5 value was selected by inspecting a histogram of the number of matches as a function of matching radius (from 0$\farcs$2 to 4$\farcs$0) at all shifted locations. We found that the median number of false matches is 96 ($\approx$12.5\% of our matched catalog) with a median offset of 1$\farcs$7.

Table~\ref{tbl:X_sources_cols} describes the 31 columns in our \CHANDRA\ source catalog. Table~\ref{tbl:X_sources} is a sample of the full catalog. This table is available online in its entirety.

\begin{deluxetable*}{@{}ll}
\tabletypesize{\scriptsize} 

\tablecaption{Overview of Columns in the {\em Chandra} Source Catalog \label{tbl:X_sources_cols}}

\tablehead{
\colhead{Column} & \colhead{Description} \\[-0.1 in]                
\setcounter{column_number}{1}
}

\startdata
1      & X-ray catalog sequence number (XID), sorted by RA. XID > 1000 are for sources in the ACIS-S3 chip. \\
2      & IAU designation. \\
3, 4   & Right ascension and declination of X-ray source (in decimal degrees) for epoch J2000.\\
5      & $\approx$80\% confidence level X-ray positional uncertainty. \\
6      & Off-axis angle $\theta$ of X-ray source (i.e., distance to average aimpoint of  {\em Chandra} observations). \\

7-15   & Source net counts in the 0.5--7 keV, 0.5--2 keV, and 2--7 keV band and corresponding upper \& lower errors. \\
16-18  & Absorbed energy flux (full, soft, and hard bands.) \\
19     & Median photon energy (full band), corrected for background. \\
20     & Fraction of the PSF (at 1.497 keV) enclosed within extraction region. PSF fraction significantly below 90\% may\\
 &       indicate a source in a crowded region. \\
21     & Absorption-corrected 0.1-2.4 keV X-ray luminosity ---for sources with optical cluster member counterpart only. \\
22     & Standard deviation of 0.1-2.4 keV X-ray luminosity ---for sources with optical cluster member counterpart only. \\

23     & Logarithmic probability $P$ that extracted X-ray counts are solely from background. \\
24     & Photometric significance computed as net counts divided by the upper error on net counts. \\
25     & Source anomalies: (f) fractional time that source was on a detector is <0.9; (a) photometry and spectrum\\
 &       may contain > 10\% afterglow events.\\
26     & Variability characterization based on K-S statistic (full band) from the single ObsId showing the most variability:\\
 &       (a) no evidence for variability (0.05<$P_{\mathrm{KS}}$); (b) possibly variable (0.005<$P_{\mathrm{KS}}$<0.05); (c) definitely variable ($P_{\mathrm{KS}}$<0.005).\\
 &       No value is reported for sources with fewer than four net counts in the full band.\\
27     & ID of optical counterpart. \\
28, 29 & Right ascension and declination of the optical counterpart. \\
30     & Offset between X-ray source and optical counterpart. \\
31     & Notes on the source. \\[-0.1 in]
\enddata

\end{deluxetable*}

\begin{deluxetable*}{@{}rc*{12}{r}ccr@{}}
\tabletypesize{\scriptsize} 

\tablecaption{{\em Chandra} Catalog:  Basic Source Properties \label{tbl:X_sources}}

\tablehead{
\multicolumn{2}{c}{Source} &
\multicolumn{4}{c}{Position} &
\multicolumn{4}{c}{Extraction} &
\multicolumn{2}{c}{Properties} \\[-0.03 in]
\multicolumn{2}{c}{\hrulefill} &  
\multicolumn{4}{c}{\hrulefill} &
\multicolumn{4}{c}{\hrulefill} &
\multicolumn{2}{c}{\hrulefill} \\
\colhead{XID}         & \colhead{CXOU J}      & \colhead{$\alpha$ (J2000)} & \colhead{$\delta$ (J2000)} & \colhead{Error}     & \colhead{$\theta$}  &
\colhead{$C_{\mathrm{f,net}}$} & \colhead{$C_{\mathrm{s,net}}$} & \colhead{$C_{\mathrm{h,net}}$}      & \colhead{PSF Frac.}         & \colhead{Anom.}    & \colhead{Var.}      \\
\colhead{}            & \colhead{}            & \colhead{(\arcdeg)}        & \colhead{(\arcdeg)}        & \colhead{(\arcsec)} & \colhead{(\arcmin)} &
\colhead{(counts)}    & \colhead{(counts)}    & \colhead{(counts)}         & \colhead{}                 & \colhead{}          & \colhead{}          \\
\colhead{(1)}         & \colhead{(2)}         & \colhead{(3)}              & \colhead{(4)}              & \colhead{(5)}       & \colhead{(6)}       &
\colhead{(7)}         & \colhead{(10)}        & \colhead{(13)}             & \colhead{(20)}             & \colhead{(25)}      & \colhead{(26)}      \\[-0.1 in]
\setcounter{column_number}{1}
}

\startdata
1 & 055132.87$+$323928.4 & 87.886986 &  32.657895 &  5.5 &  10.9 &   25.3 &    18.6 &     6.6 &  0.89 & f. & b\\
2 & 055133.77$+$322934.0 & 87.890743 &  32.492781 &  3.4 &  10.1 &   53.0 &    21.4 &    31.6 &  0.90 & .. & a\\
3 & 055134.32$+$324019.9 & 87.893001 &  32.672222 &  4.2 &  11.1 &   66.8 &    51.1 &    15.7 &  0.91 & .. & c\\
4 & 055136.34$+$323835.8 & 87.901458 &  32.643302 &  1.9 &   9.8 &  247.8 &    52.6 &   195.2 &  0.90 & .. & b\\
5 & 055136.98$+$324011.7 & 87.904097 &  32.669942 &  3.8 &  10.6 &   61.2 &    32.9 &    28.3 &  0.91 & f. & a\\[-0.1 in]
\enddata

\tablecomments{This table is available in its entirety in the electronic edition of the \apj. The first five rows with some data columns are shown here for guidance regarding its form and content. Table~\ref{tbl:X_sources_cols} describes all the columns in this table.}

\end{deluxetable*}

\section{Optical Observations}\label{optical}
\subsection{Description of the Optical Surveys}\label{optsurveys}
ME08 selected images from the survey taken with the 1.0-m telescope at the Mt.~Lemmon Optical Astronomy Observatory, AZ, of a 22$\farcs$2$\times$22$\farcs$2 area centered on M37, and used both the Scargle-Press and CLEAN periodogram techniques to measure variability for stars with 13 < $V$ < 20 mag. ME08 found 135 variables and reported \Prot\ for 120 of them, and the adopted \Prot\ for each star is the one from the two periodogram calculations with the least-scattered phased light curve. The typical \Prot\ uncertainty is $<$2\%. Based on ($B-V$) colors, these authors identified 28 of the objects as K stars, 60 as G stars, and 32 as F stars. For the latter, ME08 noted that their periods indicate pulsation instead of rotation, as F stars are not expected to display magnetic activity and, thus, no star spots.

HA08 used $g'r'i'$ filters on Megacam and the 6.5-m MMT telescope at the MMT Observatory, AZ, to survey a 24$\arcmin$$\times$24$\arcmin$ area around the center of M37. They obtained photometry for $>$38,000 objects in at least one band. HA08 converted the photometry of objects that had measurements in all three bands ($\approx$16,500) to $gri$ magnitudes. The typical one-sigma error is 0.01 mag for the $gri$ HA08 photometry, which covers the range $15\ \lapprox\ r\ \lapprox\ 23$ mag.

\citet{hartman08b} identified 1445 variable stars in the HA08 survey. These authors used the Lomb-Scargle, box-fitting least-squares, and phase-binning Analysis of Variance (AoV) algorithms to measure periodicity. The most likely period for each star was selected from these three measurements by eye.

Finally, HA09 identified periodic variability due to rotation for 575 \citet{hartman08b} variable stars. For these stars, HA09 determined \Prot\ using the multiharmonic AoV algorithm of \citet{schwarzenberg96}, which fits the harmonic series 
\begin{equation*}
\tilde{r}(t)=a_0 + \sum\limits_{i=1}^N a_i \mathrm{cos}(2\pi it /P_{\mathrm{rot}} + \phi_i),
\end{equation*}
to light curves, where $a_i$ are the amplitudes and $\phi_i$ are the phases. HA09 calculated three \Prot\ using $N=1$, 2, and 3. For the 372 stars with $<$10\% differences between the three \Prot\ values, the $N$=2 \Prot\ was adopted; these are referred to as clean periods. For the other stars, the (non-clean) period quoted in \citet{hartman08b} was kept. Here the typical \Prot\ uncertainty is $<$1\%.

\subsection{Consolidating the Optical Surveys}\label{consolidation}
We consolidated photometry and \Prot\ from the surveys described above. We matched the ME08 and HA08 positions using a 10$\arcsec$ matching radius and inspected by eye multiple-match cases to determine the most likely match using available $gri$ and $BV$ photometry as guidance. The liberal matching and individual inspection was necessary because, as noted by HA09, the absolute astrometry of ME08 objects displays a widespread non-linear distortion. Of the 153 stars in the full ME08 catalog, 16 have no HA08 counterpart. We found 37 matched stars to have  ME08 and HA09 \Prot\ measurements.\footnote{HA09 listed 22 matches to ME08 with clean \Prot, but we find that their match between HA09 ID 424 and ME08 ID 3208 appears to be a mis-identification based on their available photometry.} For consistency, we adopted the HA09 period for all 37. 

We used $gri$ photometry for our analysis. For ME08 stars, we adopted the $gri$ magnitudes of their HA08 counterparts. Of the 16 ME08 stars with no HA08 counterpart, we found three with available SDSS photometry. For the remaining 13 ME08 stars, we derived $gri$ photometry from the $V$ and $(B-V)$ reported by ME08 using the transformations derived by \citet{jester2005} and Lupton (2005).\footnote{See \url{http://www.sdss.org/dr12/algorithms/sdssUBVRITransform}} 

Our consolidated optical catalog includes 16,584 objects in the M37 field with $gri$ photometry. Of these, 657 stars have a \Prot\ measurement. Table~\ref{tbl:opt_sources_cols} summarizes the 25 columns in our catalog of optical objects, a preview of which is presented in Table~\ref{tbl:opt_sources}. This table, which includes all objects from our optical catalog with either a \Prot, a non-zero probability of being a cluster member (see discussion below), or an X-ray counterpart (see Section~\ref{matching}), is available online in its entirety.

\begin{deluxetable}{@{}ll}
\tabletypesize{\scriptsize} 

\tablecaption{Overview of Columns in Catalog of Optical Objects \label{tbl:opt_sources_cols}}

\tablehead{
\colhead{Col.} & \colhead{Description} \\[-0.1 in]                
\setcounter{column_number}{1}
}

\startdata
1      & Source ID from HA08(>10000), ME08 (<10000),\\
       & or literature. \\
2, 3   & Right ascension and declination of object (J2000). \\

4-9   & $gri$ magnitudes and their standard deviations. \\
10    & Rotation period $P_{\mathrm{rot}}$ adopted for object. \\
11    & Rotation period flag: (0) non-clean HA09 period;\\
 &       (1) clean HA09 period; (2) ME08 period. \\

12-14   & Membership probabilities ($P_{\mathrm{s}}$, $P_{\mathrm{b}}$, and $P_{\mathrm{mem}}$). \\

15     & Stellar mass. \\
16     & Convective turnover time $\tau$. \\
17     & Bolometric luminosity $L_{\mathrm{bol}}$. \\
18     & Standard deviation of $L_{\mathrm{bol}}$. \\

19     & X-ray sequence number (XID) of X-ray counterpart. \\
20     & Offset between optical object and X-ray counterpart.\\
21     & ID of ME08 counterpart ---for HA08 objects only. \\
22, 23 & Right ascension and declination of ME08 counterpart.\\
 &       ---for HA08 objects only. \\
24     & Offset between HA08 object and ME08 counterpart.\\
 &       ---for HA08 objects only. \\
25     & Notes on the object. \\[-0.1 in]
\enddata

\end{deluxetable}
\vspace{0.3 in}

\begin{deluxetable*}{@{}rc*{14}{r}ccc@{}}
\tabletypesize{\scriptsize} 

\tablecaption{Catalog of Optical Objects: Basic Properties \label{tbl:opt_sources}}

\tablehead{
\multicolumn{2}{c}{Position}   &
\multicolumn{5}{c}{Properties} &
\multicolumn{2}{c}{Membership} &
\multicolumn{3}{c}{Derived Quantities} &
\multicolumn{2}{c}{Matched Sources}    \\[-0.03 in]
\multicolumn{2}{c}{\hrulefill} &  
\multicolumn{5}{c}{\hrulefill} &
\multicolumn{2}{c}{\hrulefill} &
\multicolumn{3}{c}{\hrulefill} &
\multicolumn{2}{c}{\hrulefill} \\
\colhead{$\alpha$ (J2000)}        & \colhead{$\delta$ (J2000)}      & \colhead{$g$}                   & \colhead{$r$}  & \colhead{$i$}    & \colhead{$P_{\mathrm{rot}}$} &
\colhead{Flag} & \colhead{$P_{\mathrm{s}}$} & \colhead{$P_{\mathrm{b}}$} & \colhead{Mass} & \colhead{$\tau$} & \colhead{$L_{\mathrm{bol}}$}  &
\colhead{XID}                   & \colhead{XID offset} \\  
\colhead{(\arcdeg)} & \colhead{(\arcdeg)} & \colhead{(mag)} & \colhead{(mag)}         & \colhead{(mag)}  & \colhead{(days)}                                       &
\colhead{}          & \colhead{}          & \colhead{}      & \colhead{($M_{\odot}$)} & \colhead{(days)} & \colhead{(erg s$^{-1}$)} & 
\colhead{}          & \colhead{(\arcsec)} \\
\colhead{(2)} & \colhead{(3)} & \colhead{(4)}  & \colhead{(6)}  & \colhead{(8)}  & \colhead{(10)}  & \colhead{(11)}  &
\colhead{(12)} & \colhead{(13)} & \colhead{(15)} & \colhead{(16)} & \colhead{(17)} & \colhead{(19)}  & \colhead{(20)}\\[-0.1 in]
\setcounter{column_number}{1}
}
\startdata
87.824771 & 32.585669 & 19.03 & 17.96 & 17.33 & \nodata & \nodata & 0.36 & 0    & 0.78 & 25.90 & 7.80E+32 & \nodata & \nodata \\
87.887083 & 32.658222 & 17.48 & 16.60 & 16.12 & 5.39 & 2          & 0    & 0.04 & 1.01 & 17.11 & 2.34E+33 & 1  & 1.2 \\
87.958613 & 32.586522 & 16.13 & 15.51 & 15.27 & 3.92 & 1          & 0.51 & 0.04 & 1.19 & 12.96 & 6.37E+33 & 40 & 1.1 \\
87.966813 & 32.596517 & 22.42 & 20.66 & 19.43 & 1.24 & 1          & 0.40 & 0    & 0.50 & 48.35 & 1.82E+32 & 51 & 0.6 \\
88.002083 & 32.613389 & 15.08 & 14.15 & 13.72 & 1.89 & 2          & 0    & 0    & 1.57 & \nodata & 2.33E+34 & 108 & 0.1 \\[-0.1 in]
\enddata
\tablecomments{Only optical objects with either $P_{\mathrm{rot}}$, $P_{\mathrm{mem}}>0$, or an X-ray counterpart are included in this list. This table is available in its entirety in the electronic edition of the \apj. Five rows with some data columns are shown here for guidance regarding its form and content. Table~\ref{tbl:opt_sources_cols} describes all the columns in this table.
}
\end{deluxetable*}

\subsection{Membership Determination}\label{membership}
ME08 and HA08 relied solely on distance from a fiducial isochrone in a color-magnitude diagram (CMD) to determine M37's membership. Other studies have used radial velocity or proper motion measurements to determine membership \citep[e.g.,][]{dias2002, nilakshi2002, mermilliod2008, frinchaboy2008}, but these surveys have very few measurements and typically only cover the blue end of the MS or the cluster's red giants. We therefore used the $gi$ photometry of HA08 for $\approx$16,800 objects with $-0.5$ < $(g-i)$ < 4 and combined it with the stars' distances from M37's center to determine  membership probabilities (\Pmem).

\begin{figure}
\includegraphics{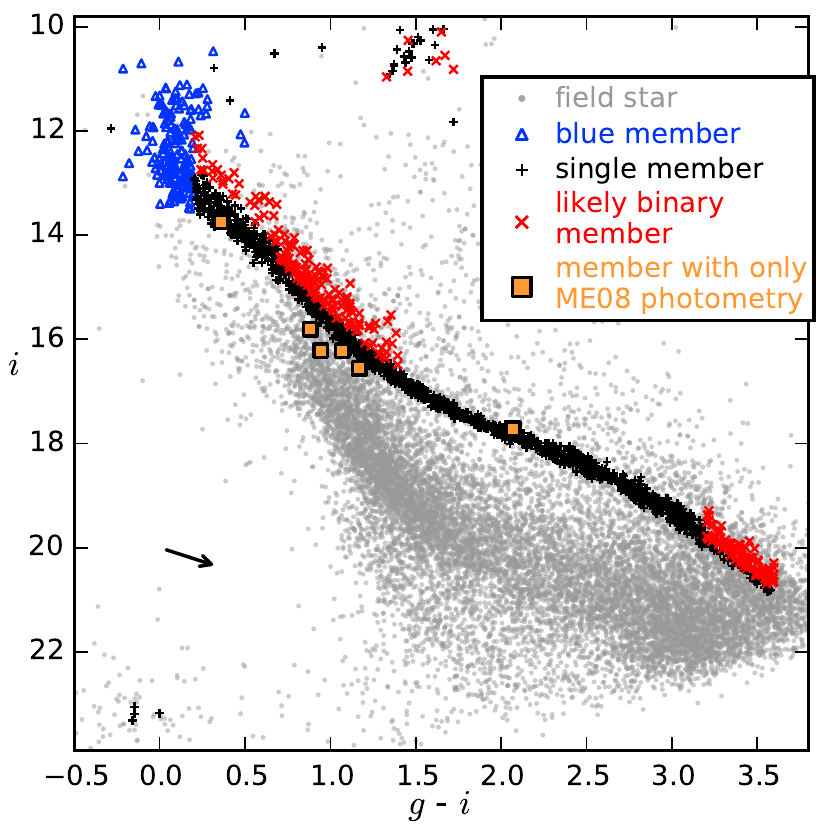}
\caption{CMD for all optical objects in the M37 field. Objects with \Pmem\ $\geq0.2$ are likely cluster members (blue, black, red, and orange symbols); objects with \Pmem\ $<0.2$ are likely field stars (gray points). Members with $P_{\mathrm{b}}$ > $P_{\mathrm{s}}$ are flagged as likely binary members (red crosses). Blue members (i.e., those beyond the MS turn-off) have no single/binary flag (blue triangles). Orange-filled squares indicate ME08 stars without HA08 counterpart classified as members. The faint blue members are white dwarfs \citep{kalirai2005}. The black arrow indicates the extinction vector.}
\label{fig:CMD}
\end{figure}

The full description of our \Pmem\ calculations is in Appendix~\ref{membershipcalc}. Stars with $0.2<(g-i)<3.6$ were assigned a probability of being a single star ($P_{\mathrm{s}}$), a binary star ($P_{\mathrm{b}}$), or a field (non-member) star ($P_{\mathrm{f}}$), so that $P_{\mathrm{s}} + P_{\mathrm{b}} + P_{\mathrm{f}} = 1$ for every star. We considered stars with \Pmem\ $\equiv P_{\mathrm{s}}$ + $P_{\mathrm{b}}\ge0.2$ to be candidate cluster members. Members with $P_{\mathrm{b}}$ > $P_{\mathrm{s}}$ were labeled as likely binaries.  The low \Pmem\ threshold we used allows for a non-negligible number of contaminants into our cluster sample. However, since we expect most field stars to be much older and, thus, inactive or significantly fainter in X rays, the contamination in our rotation-activity analysis of cluster stars will be minimal. In Figure~\ref{fig:CMD}, we show the ($i$,~$g-i$) CMD and we highlight single members, likely binary members,  and field stars.

We inspected objects with $0.1\leq P_{\rm mem} <0.2$ that also have an X-ray counterpart with a hardness ratio $<$$-0.3$ (see Sections~\ref{matching} and \ref{HR},) as these could potentially be stars that are younger than the average contaminating field star, and hence potential cluster members. We found 12 objects that met these criteria, and we assigned them a \Pmem\ value of  $999$. We use this \Pmem\ value as a flag to highlight those stars that we consider to be cluster members based on additional criteria.

For stars bluer than $(g-i) = 0.2$, we defined a separate CMD locus and calculated \Pmem\ for those based only on distance from the cluster center and without distinguishing between $P_{\mathrm{s}}$ and $P_{\mathrm{b}}$. This is because the MS is almost vertical in $(g-i)$ for these stars, so that our standard approach cannot be applied. Objects considered cluster members based on these criteria are labeled as blue members. We show these as blue triangles in Figure~\ref{fig:CMD}.

We did not apply our \Pmem\ calculation to stars without HA08 $gri$ photometry, namely, the 16 ME08 stars without an HA08 counterpart. Instead, we determined their membership by visually inspecting their location on the ($i$,~$g-i$) CMD using their $BV$ photometry transformed into $gri$ (see Section~\ref{consolidation}.) We also considered whether they matched to a soft (hardness ratio <$-0.3$) \CHANDRA\ source. Six of these stars lie near or inside the locus of MS stars on the CMD and/or match a soft \CHANDRA\ source. We assigned to these six either $P_{\mathrm{s}}=999$ or $P_{\mathrm{b}}=999$, based on their binary flag in ME08, and we indicate them with orange-filled squares in Figure~\ref{fig:CMD}.

We examined the literature for cluster membership information for stars beyond the MS turn-off. \citet{dias2002} identified 40 cluster members near the turn-off point based on proper motion measurements, but \citet{frinchaboy2008} used radial velocity measurements to show that only three are M37 members. We found an HA08 match for one of these stars in our optical catalog. For the other two, we converted the available $BV$ photometry to $gri$ using the transformations described in Section~\ref{consolidation}, assigned them \Pmem\ $=999$, and added them to our optical catalog.

We searched for surveys identifying white dwarfs and red giants in the field of M37. Using a 5$''$ tolerance radius, four cluster white dwarfs in \citet{kalirai2005} and 31 cluster red giants in \citet{mermilliod2008} matched objects in our optical catalog; we assigned all these matched objects either $P_{\mathrm{s}}=999$ or $P_{\mathrm{b}}=999$, based on their binary flag in the literature.

Lastly, we matched the cluster red giants in \citet{mermilliod2008} and massive cluster stars in \citet{nilakshi2002} to our \CHANDRA\ catalog. We found that one red giant and two massive stars matched X-ray sources that had no HA08 or ME08 counterpart (see Section~\ref{matching}). For these three cluster stars, we converted their $BV$ photometry to $gri$ using the transformations described in Section~\ref{consolidation}, assigned them either $P_{\mathrm{s}}=999$ or $P_{\mathrm{b}}=999$, based on their binary flag in the literature, and added them to our optical catalog.

Our final membership catalog contains 1699 candidate M37 members, of which 265 are likely binaries. Twenty likely binary and 406 single members have \Prot\ measurements. Table~\ref{tbl:opt_sources} includes all optical objects to which we assigned a non-zero \Pmem\ value.

We note that there is a segregation of binary candidates toward the blue and red ends of the MS. That most stars at the red end of the MS are likely binaries may be a sign of mass segregation and tidal evaporation of the low-mass single stars, or it could be a sign of increased photometric uncertainty. Such mass segregation and preferential loss of low-mass single members has been observed in many other clusters, including Praesepe \citep{pinfield2003} and the Pleiades \citep{lodieu2007}, where the binary fraction increases in the lowest-mass bins in a way not seen for cool field stars \citep[e.g.,][]{duchene2013}. Furthermore, the lack of binaries at 1.4 < ($g-i$) < 3.2 suggests that the binary sequence is not sufficiently well detected to be fit as a distinct component by our algorithm; the small number of over-luminous binary systems is likely being incorporated into the fit for the field population, indicating one possible source of incompleteness. 

\section{Analysis}\label{analysis}
\subsection{Stellar Properties Derived from Optical Photometry}\label{quantities-opt}

\subsubsection{Stellar Masses and the 500-Myr Mass-Period Distribution}\label{masses}
To estimate masses for all the stars in our optical catalog, we used the mass-absolute $r$ magnitude ($M_r$) relation of \citet{adam2007}. These authors compiled a set of empirical observations from the literature to generate a complete list of derived masses and spectral energy distributions (SEDs) for B8-L0 stars, calibrating spectral types with stars in the 600-Myr-old Praesepe cluster \citep{adam2007}.

To obtain $M_r$ we estimated the total absorption in $r$ ($A_r$) using the extinction tables by \citet{schlafly2011} assuming $R_V=3.1$ and adopting a reddening of $E$($B-V$) = 0.227$\pm$0.038 (HA08). We then adopted HA08's distance to M37 of 1490$\pm$120 pc. Finally, we linearly interpolated between the \citet{adam2007} $M_r$ values to obtain masses.

\begin{figure}
\includegraphics{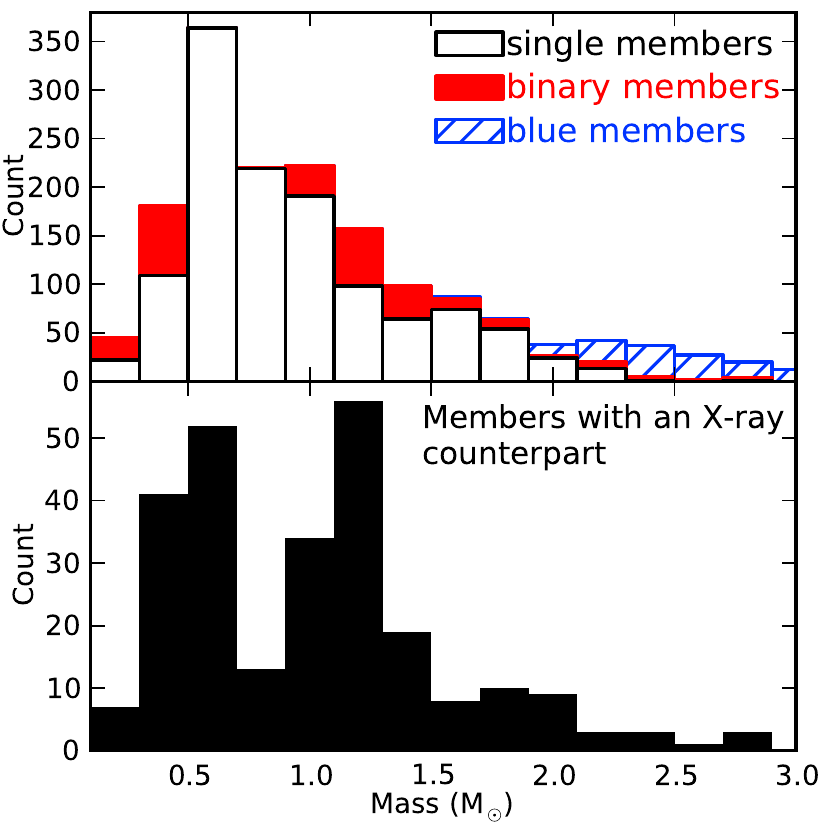}
\caption{\textit{Top}: Stacked histogram of stellar masses for M37 members. Likely binary members are indicated in red and blue members (which have no single/binary flag), in hashed blue. \textit{Bottom}: Histogram of stellar masses for M37 members with an X-ray counterpart in our \CHANDRA\ catalog. We do not derive the masses of the red giants and white dwarfs in our optical catalog and these are excluded from these histograms.}
\vspace{0.05 in}
\label{fig:masshist}
\end{figure}

\begin{figure}
\includegraphics{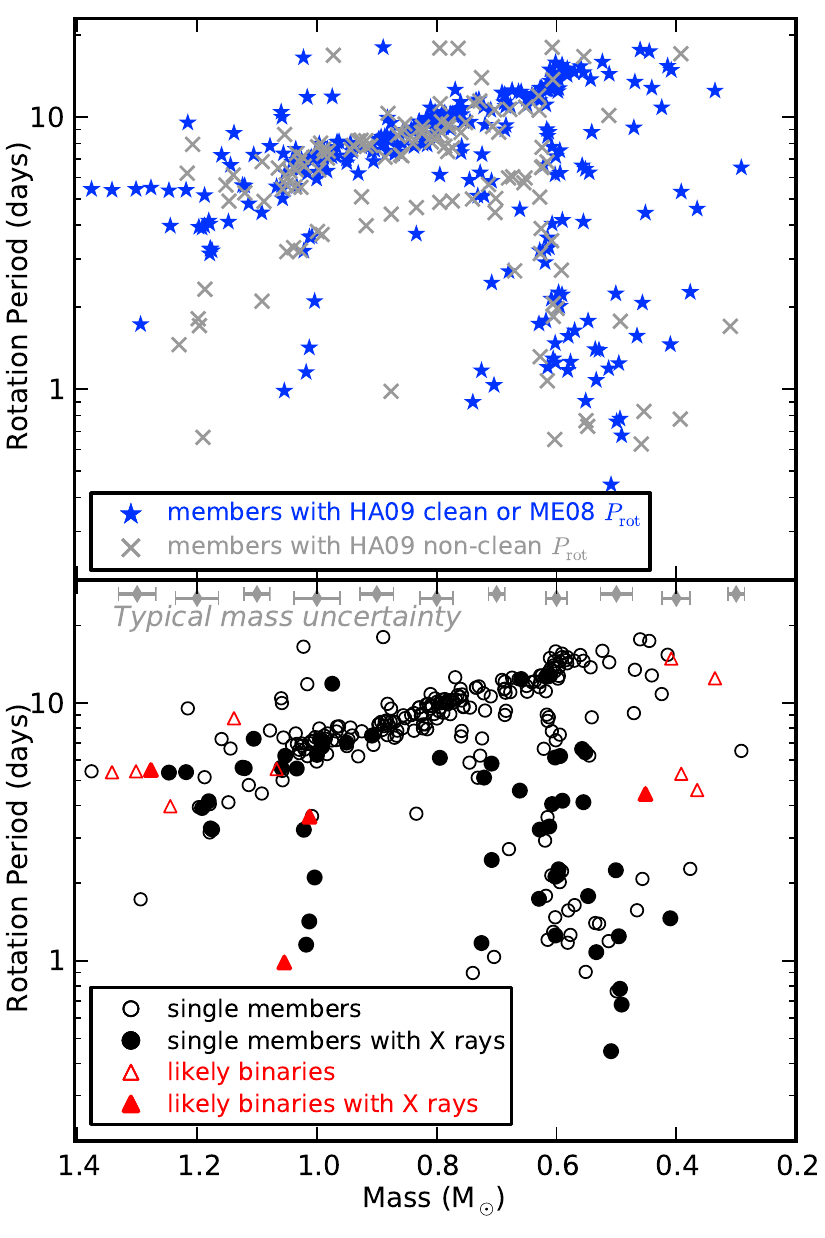}
\caption{\textit{Top}: \Prot\ versus mass for M37 members. Our consolidated optical catalog has 426 cluster members with \Prot: 296 have either HA09 clean \Prot\ or ME08 \Prot\ (blue stars), and 130 have HA09 non-clean \Prot\ (gray crosses). A well-defined, slow-rotating sequence runs from $M \approx 1.2$ \Msun, \Prot\ $\approx$ 3 d, to $M \approx 0.4$ \Msun, \Prot\ $\approx$ 20 d. This sequence breaks down around a mass of $\approx$0.7 \Msun; lower-mass stars include both fast and slow rotators. \textit{Bottom}: Same as the top panel, but for the 296 members with HA09 clean \Prot\ or ME08 \Prot. Single members (283) are indicated with black circles, and likely binary members (13) with red triangles. Stars with X-ray counterparts are shown as filled symbols (69 single members and seven likely binaries). Typical mass uncertainties, derived from photometric and distance uncertainties, are indicated along the top with error bars around gray diamonds.}
\label{fig:mass_per}
\end{figure}

Our optical catalog includes stars ranging from 0.16 to 3.8 $\Msun$. The top histogram in Figure \ref{fig:masshist} shows the derived masses for cluster members, and the bottom histogram includes only those members with X-ray counterparts in our \CHANDRA\ catalog. We find X-ray emitting cluster members covering the entire surveyed mass range. We derived mass uncertainties from photometric and distance uncertainties; we indicate typical uncertainty values with gray diamonds and error bars in the bottom panel of Figure~\ref{fig:mass_per}, the mass-period distribution for cluster stars. We distinguish between members with HA09 clean or ME08 \Prot\ and those with HA09 non-clean \Prot\ in the upper panel (see Section~\ref{optsurveys}), and between single and binary members with HA09 clean or ME08 \Prot\ in the lower panel.

There is a well-defined sequence of slow rotators in Figure~\ref{fig:mass_per} that runs from $M \approx 1.2\ M_{\odot}$, \Prot\ $\approx$ 3 d, to $M \approx 0.4\ M_{\odot}$, \Prot\ $\approx$ 20 d. Below this sequence is another distinguishable sequence composed mainly of rotators with non-clean \Prot\ that are roughly half the \Prot\ of the slow rotators. This suggests that the second sequence is composed of stars whose reported \Prot\ are a harmonic of their true \Prot. 

Although more than half of the cluster members with non-clean \Prot\ lie in or near the main mass-period sequence and therefore display the same behavior as stars with clean \Prot, we conclude that non-clean \Prot\ add unnecessary ambiguity to our results. We therefore included non-clean rotators in our analysis only if they also had a ME08 \Prot\ and the two \Prot\ measurements did not differ by more than 20\%, which is the typical vertical width of the main mass-period sequence.

The mass-period distribution for M37 indicates that the more massive cluster members have spun down to the slow-rotating sequence, where \Prot\ has a weak dependence on stellar mass. This breaks down around $\approx$0.7 \Msun; lower-mass stars span a range of 1 $\lapprox$ \Prot\ $\lapprox$ 20 d. This is consistent with expectations based on M37's slightly younger age than the Hyades and Praesepe. In the 625-Myr-old Hyades \citep{perryman1998} and 600-Myr-old Praesepe clusters, this transition occurs at $\approx$0.6 \Msun\ \citep{agueros11, Douglas14}.

Cluster members with X-ray counterparts (filled symbols in the lower panel of Figure~\ref{fig:mass_per}) cover the entire mass-period plane, suggesting that X-ray detection is not just limited to the fast rotators ($P_{\mathrm{rot}}\lesssim1$ d) and, hence, to the more active stars. Nine of the 18 fast rotators in our cluster sample have no X-ray counterpart. Eight of those nine rotators are at an off-axis angle $\theta >5\arcmin$ in the \CHANDRA\ field-of-view, beyond which the \CHANDRA\ PSF quickly degrades. Furthermore, based on their derived masses, four of them are likely to be early F stars, and the measured periods are probably due to pulsations and not to rotation.

\subsubsection{Bolometric Luminosities}\label{bolometric}
To obtain $L_{\mathrm{bol}}$ for the stars in our catalog, we once again adopted HA08's distance to M37 of 1490$\pm$120 pc. We used the SED table of \citet{adam2007} to convert the $M_r$ values obtained in Section~\ref{masses} into corresponding effective temperatures $T_{\mathrm{eff}}$. We then used $T_{\mathrm{eff}}$ to find bolometric corrections $BC_r$ using the tables collected by \citet{Girardi04}, which are tailored to the SDSS filter system. Using $M_r$ and $BC_r$, we then calculated bolometric magnitudes and luminosities.

\subsubsection{Rossby Numbers}\label{rossby}
Stellar activity increases with increasing rotation rates only up to a saturation velocity, beyond which activity remains constant. This saturation point has been observed to have a dependence on stellar mass \citep[see e.g.,][]{pizzo03}. \citet{Noyes1984} found that the Rossby number $R_o$ = $P_{\mathrm{rot}}/\tau$, where $\tau$ is the convective turnover time, removes this mass dependence, thus revealing the general relationship between stellar rotation and activity. $R_o$ serves as an indicator of the strength of a rotationally driven dynamo, which can be compared more evenly between stars of different masses.

To estimate $\tau$ for M37 members, we used the empirical stellar mass-$\tau$ relation of \citet[][hereafter W11]{wright2011}, who collected over 800 stars spanning a mass range of 0.09--1.36 $M_{\odot}$ with both \Prot\ and $L_{\rm X}$ measurements. These authors fitted the \LL-\Prot\ relation following a formulation equivalent to Equation~\ref{eq:rossby} and adopting the canonical slope $\beta=-2.0$. W11 reported $\tau$ values for ten mass bins, and we used these to fit a second-order polynomial in the mass-$\tau$ plane. The resulting mass-$\tau$ relation is
\begin{equation*}
\mathrm{log} \tau = 1.24 - 1.63\ \mathrm{log} (M/\Msun) - 0.56\ \mathrm{log}^2 (M/\Msun),
\end{equation*}
with a correlation coefficient $r$ = 0.989, and we used this relation to calculate $\tau$ for cluster members within the 0.09--1.36 $M_{\odot}$ mass range. 72\% of cluster members and 99\% of the cluster rotators fall within this mass range.

W11 also reported $\tau$ values for the same mass bins adopting $\beta=-2.7$, which is the slope they found in their analysis of a subset of 36 solar-type stars. We used the second-order polynomial that W11 fit to these $\tau$ values in the mass-$\tau$ plane to calculate $\tau$ for our cluster stars, and found that these $\tau$ values were 30$\pm$5\% smaller than those based on $\beta=-2.0$. 

Using both sets of $\tau$ values, we calculated $R_o$ and examined the \LL-$R_o$ plane for our cluster sample, performing the parametrization analysis described in Section~\ref{results} twice. We found that the two $\beta$ values we derived for our sample differed $<$3\% and were well within one standard deviation ($\sigma$) of each other. The two saturation points $R_{o,\mathrm{sat}}$ differed by 20\%, but were still within 1$\sigma$ of each other. We therefore opted to use $\tau$ values based on $\beta=-2.0$ for the rest of our analysis.

\subsection{Stellar Properties Derived from X rays}\label{quantities-xray}
The stellar magnetic field manifests itself above the photosphere and heats the plasma in the stellar corona above $10^6$ K \citep{Vaiana81}. X rays from MS stars originate in this high-temperature plasma. In such stars, the ratio of X-ray luminosity to bolometric luminosity \LL\ appears to decrease by several orders of magnitude between zero-age MS stars and solar-age stars \citep[e.g.,][]{Feigelson2004, wright2010}. The \LL\ ratio thus allows us to compare the contribution of X rays in a mass-independent way.

To estimate the unabsorbed fluxes of our \CHANDRA\ sources, we converted our soft-band fluxes using WebPIMMS\footnote{\url{http://cxc.harvard.edu/toolkit/pimms.jsp}} and an APEC model with solar abundance and $kT$ = 0.768 (log $T$ = 6.95). Given that 95\% of the X-ray sources with M37 counterparts have similar, soft X-ray spectra (with hardness ratios between $-0.4$ and $-1.0$; see Section~\ref{HR}), assuming the same plasma temperature for all is a reasonable approximation. We set the Galactic atomic neutral hydrogen column density to 1.26$\times 10^{21}$ cm$^{-2}$, derived using $E$($B-V$) = 0.227 from HA08, and obtained the fluxes for the $\ROSAT$ band (0.1--2.4 keV) in order to compare our results to those of previous studies. We then calculated $L_{\mathrm{X}}$ using a distance of 1490$\pm$120 pc (HA08). 

\subsection{Matching the Optical and X-ray Catalogs}\label{matching}
We used the $\approx$80\% confidence X-ray positional uncertainties calculated in Section~\ref{uncertainty} to match our \CHANDRA\ sources to optical objects from HA08 by using a tolerance radius 1.6 times the positional uncertainty of each X-ray source. We found that this factor provided a large number of matches without introducing too many potentially false matches. We found 540 X-ray sources with HA08 optical counterparts. Multiple matches occurred $<$15\% of the time, and we selected the closest match as the true one in such cases. Fourteen more near-matches were inspected individually and considered to be true matches.

To match our X-ray sources to ME08 objects, we used the same method as described in Section~\ref{consolidation}, where we described how a distortion in the absolute astrometry of ME08 objects led us to apply a liberal 10$\arcsec$ tolerance radius and then to inspect by eye the most likely true match using available $gri$ and $BV$ photometry as guidance. We found 37 X-ray sources with ME08 optical counterparts.

For the remaining unmatched X-ray sources, we found one red giant match in \citet{mermilliod2008} and two massive star matches in \citet{nilakshi2002} (see Section~\ref{membership}) also using a tolerance radius 1.6 times the positional uncertainty of each X-ray source. Neither of these three optical objects has a HA08 or ME08 counterpart (two of them have $V$ photometry outside the $V$ range studied by ME08).

\begin{figure}
\includegraphics{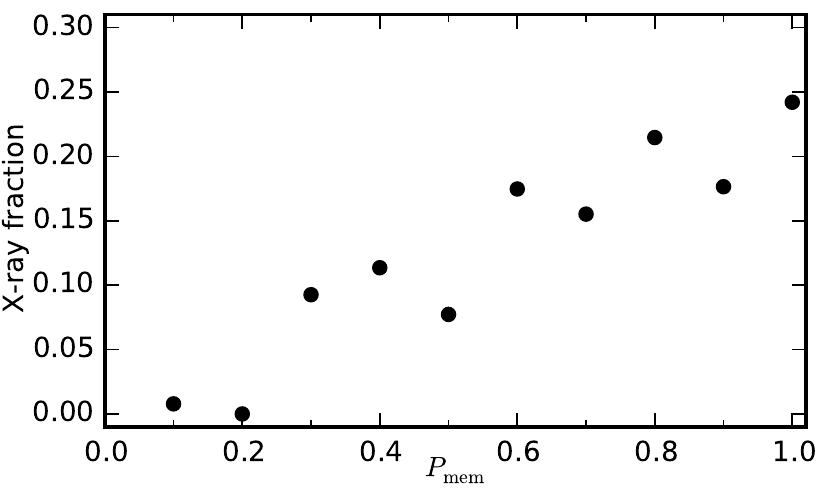}
\caption{Fraction of optical objects with X-ray counterparts per 10\%-\Pmem\ bin.}
\label{fig:Xfraction}
\end{figure}

Combining all X-ray-to-optical matches, we have 561 X-ray sources with optical counterparts, 278 of which have \Pmem\ $\geq$ 0.2: 195 single stars (69 with \Prot) and 83 binaries (7 with \Prot). Figure \ref{fig:Xfraction} shows the fraction of optical objects with X-ray counterparts as a function of \Pmem. It is close to zero for field stars and then increases linearly to some intrinsic fraction of M37 stars.

Two hundred and thirteen X-ray sources have no optical counterparts. The log of the median spurious detection probabilities $P$ for matched and unmatched X-ray sources are $-$9.5 and $-$10.6, respectively. The median net counts in the full band for matched and unmatched X-ray sources are 17.2 and 25.9, respectively. Lastly, the median photon energy for matched and unmatched X-ray sources are 1.2 and 1.95 keV, respectively.

\subsection{X-ray Hardness Ratios and Spectral Information}\label{HR}
We obtained spectral information for our X-ray sources by computing a hardness ratio ($HR$). The $HR$ is (hard counts$-$soft counts)/(hard counts$+$soft counts); we followed the definitions of the hard and soft bands described in Section~\ref{merging} and used net counts. Soft spectra are characterized by $HR < 0$, and hard spectra, by $HR > 0$. 

\begin{figure}
\includegraphics{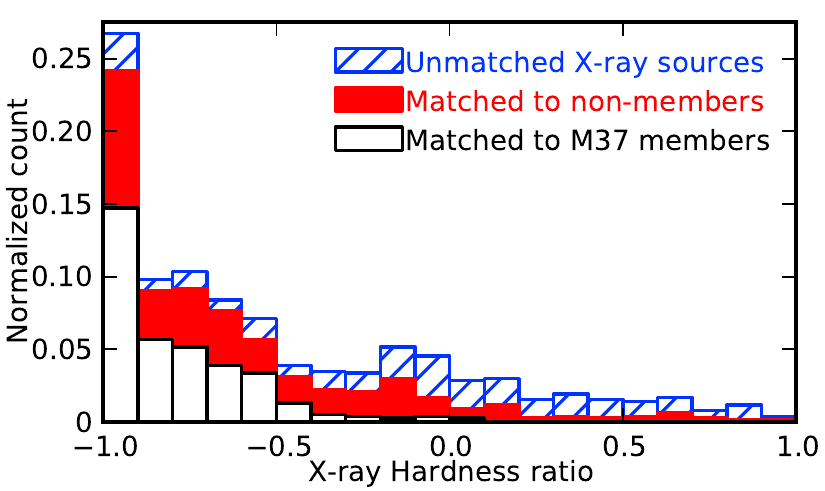}
\caption{Normalized stacked histogram of hardness ratios for X-ray sources matched to M37 members (white), X-ray sources matched to non-member objects (red), and X-ray sources with no optical counterpart (hashed blue).}
\label{fig:HR}
\vspace{0.05 in}
\end{figure}

An \textit{a posteriori} analysis of the matching done in Section~\ref{matching} revealed that 96\% of the 278 X-ray sources matched to optical cluster members have $HR \lesssim -0.3$, a strong indication of their stellar nature. Only two such sources have $HR$ > 0, and both are flagged as likely binaries. By contrast, 72\% of the 283 X-ray sources matched to optical non-members (i.e., field stars) have $HR \lesssim -0.3$, and 12\% have $HR > 0$. Furthermore, only 32\% of the 213 X-ray sources without any optical counterpart have $HR\lesssim -0.3$, and 45\% have $HR>0$. Figure~\ref{fig:HR} shows a histogram of $HR$ values for X-ray sources matched to M37 members, X-ray sources matched to non-members, and X-ray sources with no optical counterpart.

Only three X-ray sources matched to an optical cluster member have $>$150 net counts in the full band; the highest has 260. Thus, there are no cluster X-ray sources with enough counts for a meaningful analysis of their X-ray spectra or light curves. Four X-ray sources in our \CHANDRA\ catalog have  $>$1000 net counts. Two of these are matched to an optical non-member; the other two lack optical counterparts. We discuss these four sources in Appendix~\ref{spectra}.

\section{Results and Discussion}\label{results}
\begin{figure}
\includegraphics{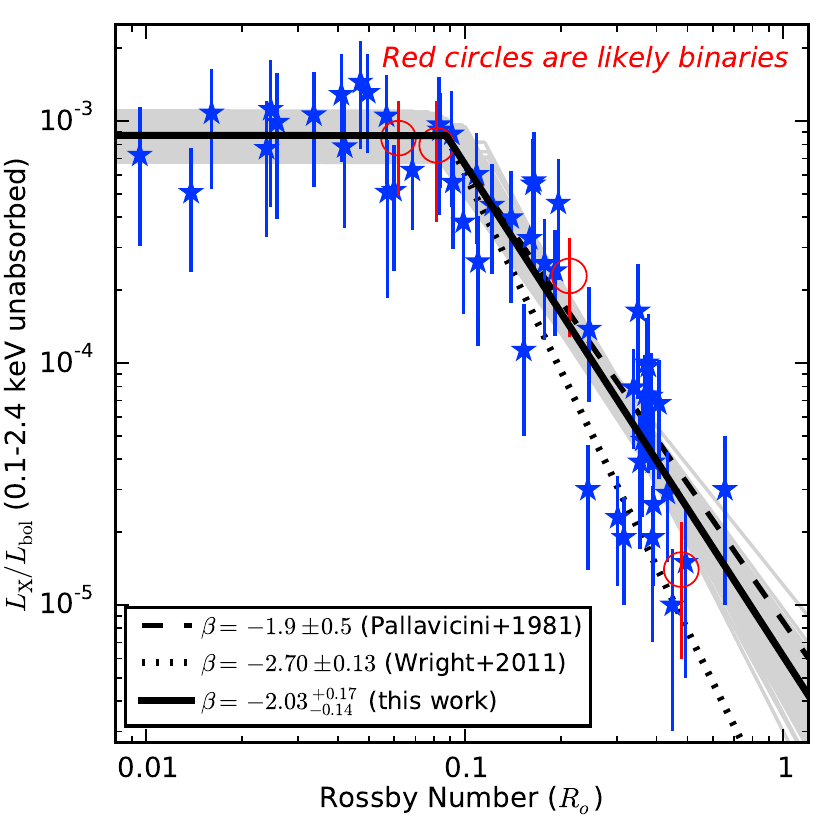}
\caption{X-ray to bolometric luminosity ratio as a function of Rossby number $R_o$ for cluster members. Likely binary members are indicated with red circles. The solid black line is the maximum $a\ posteriori$ fit from the MCMC algorithm and the gray lines are 200 random samples from the posterior probability distributions. The dashed line indicates the fit found by \citet{palla81}, and the dotted line indicates the fit found by \citet{wright2011}.}
\label{fig:rossby}
\end{figure}

Our analysis of \LL\ versus $R_o$ only included M37 members (\Pmem\ $\geq$ 0.2) that had either HA09 clean \Prot\ or that had both HA09 non-clean \Prot\ and ME08 \Prot\ and the two \Prot\ measurements differed by less than 20\%. Figure \ref{fig:rossby} shows the resulting 51 single members (blue stars) and 4 likely binary members (red circles) that are included in the following fit.  

The conventional way to parametrize the \LL--$R_o$ relationship is with a flat region connected to a power-law \citep[e.g.,][]{Randich2000b, wright2011, Douglas14}:
\begin{equation}\label{eq:rossby}
  \frac{L_{\mathrm{X}}}{L_{\mathrm{bol}}} = \left\{
  \begin{array}{l l}
    \left(\frac{L_{\mathrm{X}}}{L_{\mathrm{bol}}}\right)_{\mathrm{sat}} & \quad \textrm{if $R_o\le R_{o\mathrm{,sat}}$}\\
    C R_o^\beta & \quad \textrm{if $R_o$ > $R_{o\mathrm{,sat}}$}
  \end{array} \right.
\end{equation}
where $\left(L_{\mathrm{X}}/L_{\mathrm{bol}}\right)_{\mathrm{sat}}$ is the activity saturation level, $R_{o\mathrm{,sat}}$ is the turnover point, $\beta$ is the power-law index characterizing the unsaturated regime, and $C$ is a constant. We fit this model to the 55 members described above. 

\begin{figure}
\includegraphics{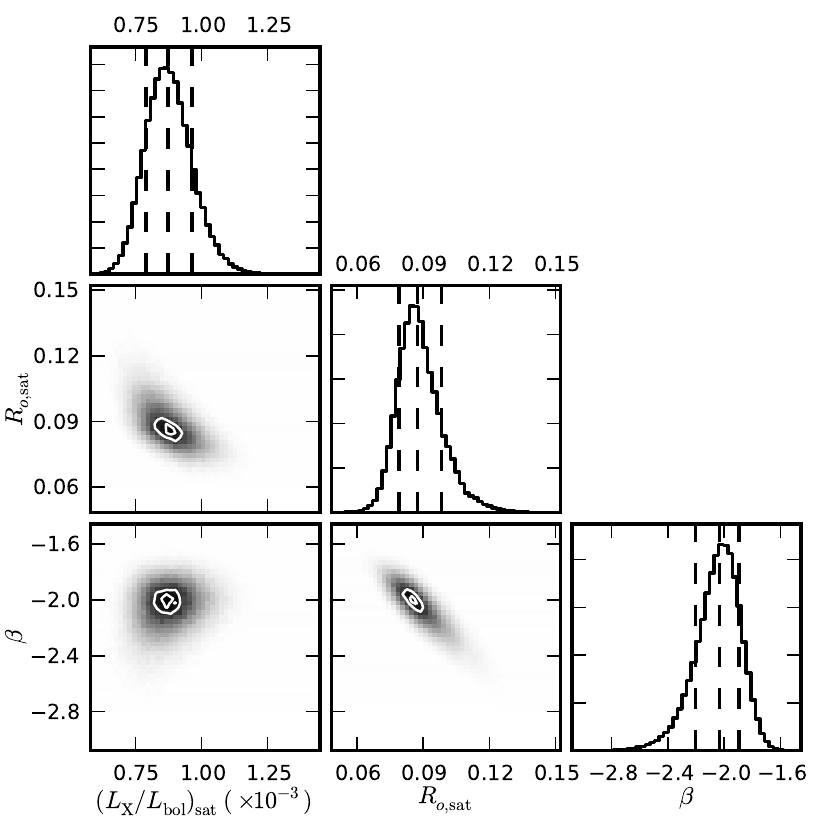}
\caption{Marginalized posterior probability distributions from the MCMC analysis using \texttt{emcee}. The parameter values of the \textit{a\ posteriori} model are the peaks of the one-dimensional distributions; the vertical dashed lines approximate the median and 68-percentiles. The two-dimensional distributions illustrate covariances between parameters; the contour lines approximate the 68- and 95-percentile of the distributions.}
\label{fig:corr}
\end{figure}

We followed a similar fitting procedure to that described in \citet{Douglas14}. We converted the model in Equation \ref{eq:rossby} into log-space for the fit, and assumed flat priors over all three parameters. We used the open-source Markov-chain Monte Carlo (MCMC) package \texttt{emcee}\footnote{\url{http://dan.iel.fm/emcee/current/}} \citep{Foreman13} to carry out the fit. It builds posterior probability distributions for each parameter (shown in Figure~\ref{fig:corr}) by performing a random walk in parameter space. 

Two hundred models drawn at random from the posterior probability distributions are shown in gray in Figure~\ref{fig:rossby}. The most probable model, i.e., the maximum $a\ posteriori$ model, is indicated by a solid black line. The resulting best-fit parameters are log(\LL)$_{\mathrm{sat}} = -3.06\pm0.04$, $R_{o\mathrm{,sat}}=0.09\pm0.01$, and $\beta=-2.03_{-0.14}^{+0.17}$. The parameter values correspond to the 50$^{\rm th}$ quantile, and the uncertainties correspond to the 16$^{\rm th}$ and the 84$^{\rm th}$ quantiles (for consistency with $1\sigma$ Gaussian uncertainties).

\subsection{The Saturated Regime}\label{saturated}
Several hypotheses aim to explain the decoupling of rotation and activity seen in the saturated regime. These include dynamo saturation \citep{Gilman83, Vilhu87}, saturation of the filling factor of active regions on the stellar surface \citep{Vilhu84}, and centrifugal stripping of the corona \citep{Jardine1999}. However, a satisfactory explanation remains elusive (see e.g., W11). 

We detect this saturation phenomenon for stars with $R_o < 0.09\pm0.01$. This turnover point is similar to $R_{o\mathrm{,sat}}= 0.13\pm0.02$ found by W11 for their $Einstein/\ROSAT/XMM$ heterogeneous sample of 824 field and cluster stars collected from the literature, and it is slightly smaller than $R_{o\mathrm{,sat}}\approx0.16$ found by \citet{Randich2000b} for her heterogeneous $\ROSAT$ sample of field and cluster stars. Our saturation level log$\left(L_{\mathrm{X}}/L_{\mathrm{bol}}\right)_{\mathrm{sat}}=-3.06\pm0.04$ agrees with log$\left(L_{\mathrm{X}}/L_{\mathrm{bol}}\right)_{\mathrm{sat}}=-3.13\pm0.22$ found by W11. It is slightly lower than log$\left(L_{\mathrm{X}}/L_{\mathrm{bol}}\right)_{\mathrm{sat}}\approx-3$ found by \citet{Randich2000b}.

\subsection{The Unsaturated Regime}\label{unsaturated}
The increase in stellar activity with increasing rotation (smaller $R_o$) is believed to indicate the heating of the corona by the stellar dynamo \citep{palla81, pizzo03}. One of the challenges of the past few decades has been to characterize empirically this relationship. \citet{palla81} first discovered this behavior in an $Einstein$ sample of 17 late-type stars of luminosity classes II to V, and found a power-law index $\beta=-1.9\pm0.5$. More recently, \citet{gudel1997} found $\beta = -2.64\pm0.12$ for a $\ROSAT/ASCA$ sample of 12 solar-type stars of ages 0.07 to 9 Gyr, and \citet{Randich2000b} found $\beta = -2.10\pm0.09$ for her $\ROSAT$ sample. Finally, W11 found $\beta = -2.18\pm0.16$ for their heterogeneous sample of cluster and field stars. W11 also analyzed a subset of 36 solar-type stars detected only with $\ROSAT$ to eliminate X-ray luminosity biases. They found $\beta = -2.70\pm0.13$ for this sample. We include the power-law relation for the unsaturated regime of both the latter sample and of the \citet{palla81} sample in Figure~\ref{fig:rossby}.

For our sample of unsaturated, co-eval M37 stars, which cover a mass range of 0.55--1.28 \Msun, the index $\beta=-2.03_{-0.14}^{+0.17}$ agrees with that by \citet{palla81} $\beta=-1.9\pm0.5$. It differs from the W11 result for the unbiased sample of solar-type stars by 3$\sigma$. That $\beta$ found for our sample of $\approx$500 Myr stars does not agree with some of the $\beta$ values in the literature suggests that previous studies that include stars with a range of ages may be missing interesting age-dependent effects in the behavior of stellar activity. It may also be evidence that the unsaturated regime is better described by a broken power law instead of a single power law \citep[e.g.,][]{petit2008,donati2009,saar1999}. Unfortunately, our sample of M37 does not extend to large enough $R_o$ values to fully test this idea.

Did our \CHANDRA\ observation or the \Prot\ surveys of ME08 and HA09 bias our sample toward those stars with higher $L_{\rm X}$ and/or faster \Prot? We may not have identified the faintest X-ray sources in the cluster because of their very low counts and/or large $\theta$, making their detection much harder. Likewise, ME08 and HA09 may not have measured \Prot\ for the faintest cluster members, thus excluding members with possibly the longest \Prot.

Our X-ray catalog contains 18 sources, 15 of which are cluster members, with log(\LL) $\leq-6.0$, while our optical catalog contains only six objects, three of which are members, with $R_o\geq 1.0$. Indeed, the light curves used by both ME08 and HA09 span only 30 days, suggesting that lack of slow rotators may be the limiting factor in determining the behavior of unsaturated stars at the high-$R_o$ end.

Still, our sample of $\approx$500 Myr M37 members with both \Prot\ and X-ray measurements spans a wide set of stellar properties, as it covers masses ranging from 0.4 to 1.3 \Msun, \Prot\ ranging from 0.4 to 12.8 d, and $L_{\rm X}$ ranging from 10$^{28.4}$ to $10^{30.5}$ \ergs. Our results make M37 a new benchmark open cluster for calibrating the rotation--activity relation at ages of $\approx$500 Myr. 

\subsection{The Dependence of X-ray Activity on Age}\label{LX_age}
\begin{figure}
\includegraphics{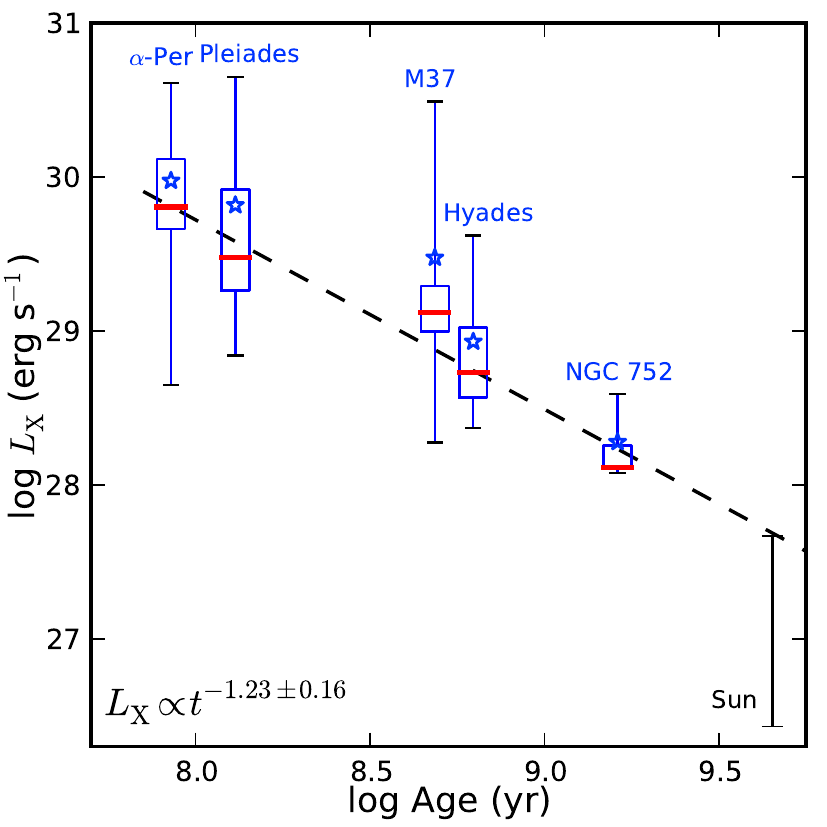}
\caption{X-ray luminosity of solar-like members (mass range of 0.8--1.2 \Msun) of five open clusters as a function of age. The adopted ages are 85$\pm$10 Myr for $\alpha$-Per, 130$\pm$20 Myr for the Pleiades, 485$\pm$28 Myr for M37, 625$\pm$50 Myr for the Hyades, and 1.6$\pm$0.2 Gyr for NGC 752. The data includes 62 $\alpha$-Per stars, 62 Pleiades stars, 64 M37 stars, 28 Hyades stars, and 5 NGC 752 stars. For each cluster, the box extends from the lower to the upper quartile, and the whiskers cover the entire $L_{\rm X}$ range. The median $L_{\rm X}$ for each cluster is indicated by a red line inside the box, and the mean by a blue star. We indicate the minimum and maximum of the Sun \citep{peres2000} with black whiskers at 4.5 Gyr. A linear regression analysis in log space (excluding the Sun) suggests a relationship between X-ray luminosity and age of $L_{\mathrm{X}}\propto t^{-1.23\pm0.16}$, and we indicate it with a dashed line.}
\label{fig:LX_t}
\vspace{0.05 in}
\end{figure}

We compared $L_{\mathrm{X}}$ for 64 M37 solar analogs (masses 0.8--1.2 \Msun) with X-ray detections of 62 stars in Alpha Persei \citep[$\alpha$-Per,][]{randich1996, Prosser1996, Prosser1998, Prosser1998b, pillitteri2013}, 62 Pleiads \citep{micela1990, stauffer1994, micela1996, micela1999, briggs2003}, 28 Hyads \citep{stern1994, stern1995}, and five stars in NGC 752 \citep{giardino2008} within the same mass range. The adopted ages are 85$\pm$10 Myr for $\alpha$-Per \citep{lodieu2012}, 130$\pm$20 Myr for the Pleiades \citep{barrado2004}, 485$\pm$28 Myr for M37 (HA08), 625$\pm$50 Myr for the Hyades \citep{perryman1998}, and 1.6$\pm$0.2 Gyr for NGC 752 (Ag\"ueros et al.~in prep.). Figure~\ref{fig:LX_t} shows the data for these five clusters, where the boxes extend from the lower to the upper quartile, the whiskers extend to cover the entire range, the red line inside the boxes indicates the median, and the blue stars indicate the mean. The maximum and minimum $L_{\mathrm{X}}$ of the Sun \citep{peres2000} is indicated by black whiskers at 4.5 Gyr.

We performed a linear regression analysis in log space using the median $L_{\mathrm{X}}$ values of the five clusters to determine the dependence of $L_{\mathrm{X}}$ on age. We found that $L_{\mathrm{X}}\propto t^{-1.23\pm0.16}$, with a correlation coefficient $r=-0.98$. This result, which corresponds to the black dashed line in Figure~\ref{fig:LX_t}, is close to the expected power-law exponent value of $-1$ for stars with ages $\gapprox$100 Myr. It is within 1$\sigma$ of the power-law exponent $-1.5^{+0.3}_{-0.2}$ found by \citet{maggio1987} using a set of Pleiades, Hyades, and Orion Ic solar-like stars and their observations of nearby solar analogs. Similarly, \citet{gudel1997} and \citet{giardino2008} found an exponent of $-1.5$. Plugging our $\beta=-2.03_{-0.14}^{+0.17}$ into the relation $L_{\mathrm{X}}\propto t^{-\alpha\beta}$, we find that $\alpha$ must lie in the range 0.49--0.75, with $\alpha=0.60$ as the most likely value. This agrees very well with the relation $v_{\mathrm{rot}} \propto t^{-0.5}$ for solar-like stars found by \citet{skumanich72} and \citet{meibom2011}.

\section{Summary}\label{conclusion}
We have assembled an extensive sample of stars with X-ray and \Prot\ measurements in the M37 open cluster to study the stellar rotation-activity relation at $\approx$500 Myr. We observed the center of the cluster with \CHANDRA\ and detected $\approx$770 high-confidence X-ray sources. We also consolidated two optical surveys of M37 with \Prot\ measurements \citep{messina08a, hartman09}, obtaining a merged sample of $\approx$650 rotators. We calculated new membership probabilities for optically detected objects in the M37 field, and complemented the resulting catalog of cluster members with massive stars, red giants, and white dwarfs identified as members in the literature. By matching our X-ray and optical catalogs, we generated a sample of 76 solar- and late-type members of M37 with X-ray and \Prot\ measurements.

We used soft-band X-ray fluxes extracted from our \CHANDRA\ observations and the cluster distance of $\approx$1.5 kpc to derive $L_{\mathrm{X}}$ for all X-ray sources. We then used $gri$ photometry and reddening values from \citet{hartman08a} to derive stellar masses and $L_{\mathrm{bol}}$ for all cluster members. Finally, we calculated $\tau$ values using the \cite{wright2011} empirical model as a function of mass, thereby obtaining $R_o$ for our stars.

We then examined the $\approx$500 Myr mass-rotation and rotation-activity relations. M37's mass-period distribution indicates that the cluster's more massive stars have spun down to a single-valued relation. There is a well-defined sequence of slow rotators that runs from $M \approx 1.2\ M_{\odot}$, \Prot\ $\approx$ 3 d, to $M \approx 0.4\ M_{\odot}$, \Prot\ $\approx$ 20 d. This single-valued relation breaks down around a mass of 0.7 \Msun; lower-mass stars span a range from fast (\Prot\ $\approx 1$ d) to slow ($\approx$20 d) rotators. This is consistent with expectations based on M37's slightly younger age than the Hyades and Praesepe. In these $\approx$600-Myr-old clusters, this transition occurs at $\approx$0.6 \Msun\ \citep{agueros11, Douglas14}.

In examining the dependence of \LL\ on $R_o$, we identified a saturated and an unsaturated regime. All rotators with $R_o<0.09\pm0.01$ (a turnover point consistent with that found in the literature for other rotation-activity samples) are saturated, and converge to an activity level of log(\LL)$=-3.06\pm0.04$, close to the canonical value of $-3$ and similar to recent values found for heterogeneous samples of field and cluster stars.

In the unsaturated regime, faster rotators (smaller $R_o$) have increasing levels of coronal activity (larger \LL) following a power law of index $\beta=-2.03^{+0.17}_{-0.14}$, in agreement with the $\beta=-1.9\pm0.5$ found by \citet{palla81}. Interestingly, however, this $\beta$ differs from the \citet{wright2011} result for their unbiased sample of solar-type stars by 3$\sigma$. This suggests that previous studies that include stars with a range of ages may be missing interesting age-dependent effects in the behavior of stellar activity, or that a broken power law may describe better the unsaturated dependence of activity on rotation.

A comparison of solar-type X-ray emitters in M37 and similar stars in the $\alpha$-Per, Pleiades, Hyades, and NGC 752 open clusters, covering an age range of 0.85--1.6 Gyr, shows that $L_{\mathrm{X}}\propto t^{-1.23\pm0.16}$, suggesting that the surface rotational velocity in solar-type stars declines as $v_{\mathrm{rot}} \propto t^{-0.60}$, in agreement with previous studies \citet[e.g.,][]{skumanich72, meibom2011}.

By probing the stellar rotation-activity relation with our homogeneous, co-eval sample of M37 members, we are helping to calibrate empirically the ARAR at $\approx$500 Myr. More importantly, the wide coverage in stellar masses, rotation periods, and X-ray luminosities of our sample of cluster members puts M37 in a position to become a new benchmark middle-aged open cluster.

\acknowledgments We thank the referee for a careful reading of the paper. We thank Patrick Broos for his help with the AE package and Maureen van den Berg for her help with obtaining unabsorbed X-ray fluxes. Support for this work was provided by NASA through \CHANDRA\ Award Number G02-13025A issued by the Chandra X-ray Observatory Center, which is operated by the Smithsonian Astrophysical Observatory for and on behalf of NASA under contract NAS8-03060. M.A.A.\ acknowledges support provided by the NSF through grant AST-1255419.

\setlength{\baselineskip}{0.6\baselineskip}
\bibliography{references}
\setlength{\baselineskip}{1.667\baselineskip}

\appendix

\section{Calculation of membership probabilities}\label{membershipcalc}
ME08 and HA08 relied solely on distance from a fiducial isochrone in a CMD to determine membership for M37. Other studies have used radial velocity or proper motion measurements to determine membership \citep[e.g.,][]{dias2002, nilakshi2002, mermilliod2008, frinchaboy2008}, but these surveys have very few measurements and typically only cover the blue end of the MS or the cluster's red giants. Since kinematic measurements of objects in the field of M37 are scarce, we decided to use the spatial distribution of stars, in addition to their location in a CMD, to determine membership probabilities. 

\subsection{Description of the Method}\label{mem_method}
\begin{figure}
\centerline{\includegraphics{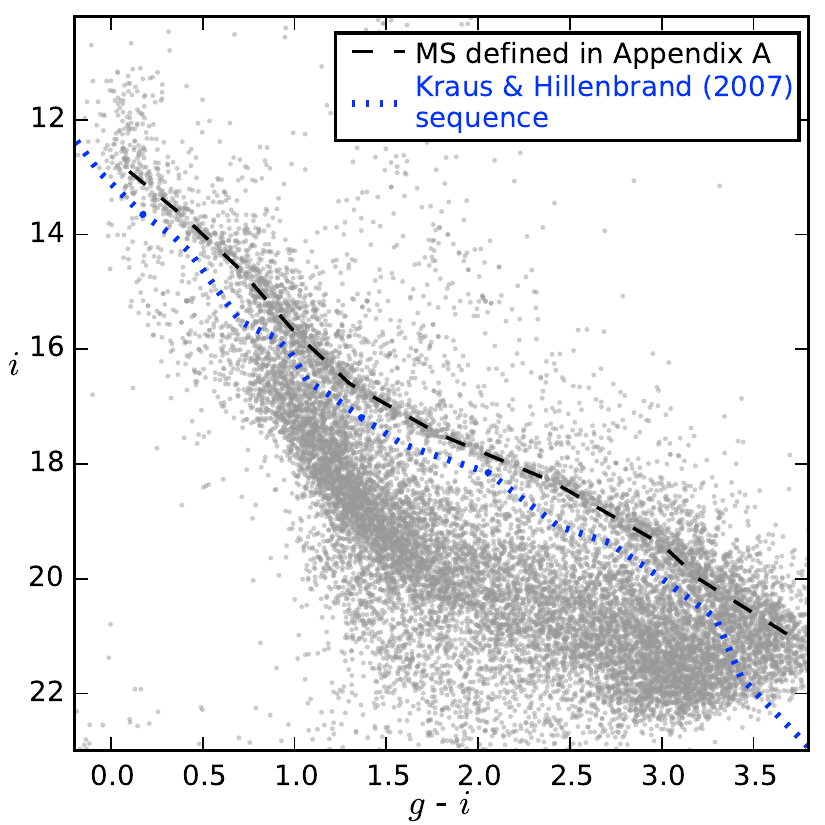}}
\caption{CMD for all optical objects in the M37 field. The dashed line indicates the MS defined in this paper by fitting a polynomial to a subset of stars close to the cluster center that made the MS visually obvious. The blue dotted line indicates the color-magnitude sequence derived by \citet{adam2007} for Praesepe.}
\label{fig:CMDplain}
\vspace{0.1 in}
\end{figure}

First, we binned all the stars in the M37 field by $(g-i)$ color, which allows us to keep both single and binary cluster stars within the same bins. We then defined a MS in the ($i$, $g-i$) CMD by fitting a polynomial to a subset of stars close to the cluster center that made the MS visually obvious. Figure~\ref{fig:CMDplain} shows the full set of optical objects in the M37 field, and we indicate with a dashed line the MS we defined. We also include the color-magnitude relation derived by \citet{adam2007} for Praesepe members. The slight discrepancy between the two sequences is expected due to the differences in filter systems and in metallicity, age, and reddening. Next, we built probability distribution functions (PDFs) for the single, binary, and field star populations. The two quantities that we used for fitting in each ($g-i$) bin were the radius $r$ from the cluster center and height $m=i-i_{\mathrm{MS}}$ above or below the MS in the ($i$, $g-i$) CMD. By thinking in terms of height above/below the sequence (i.e., fitting and subtracting the MS), we worked in terms of a quantity for which the distribution of contaminants is roughly constant across the bin at a given height/depth. Figure~\ref{fig:memMS} shows the distribution of $m$ for stars in the M37 field in terms of $(g-i)$ color and radius from the cluster center. Stars are color-coded using the membership probabilities obtained after performing the fit described next.

The parameters that we fitted in the PDFs were:
\begin{enumerate}
\item Three normalization constants: the number of single cluster stars ($N_{\mathrm{s}}$), of binary cluster stars ($N_{\mathrm{b}}$), and of field stars ($N_{\mathrm{f}}$). Clearly, $N_{\mathrm{s}}$+$N_{\mathrm{b}}$+$N_{\mathrm{f}}$=$N_{\mathrm{tot}}$. Since $N_{\mathrm{tot}}$ is set, there were really only two parameters being fitted.

\item One spatial distribution parameter: the exponential scale radius of the cluster $r_0$. We assumed a flat spatial distribution of field stars, which is defined by the size of the field and $N_{\mathrm{f}}$ and, hence, does not introduce any additional fitting parameters.

\item Two parameters setting the intrinsic width of the MS in terms of $\sigma$ of $m$: one around the single star sequence ($\sigma_{\mathrm{s}}$) and one around the binary star sequence ($\sigma_{\mathrm{b}}$).

\item Two parameters setting the median height of the MS: one for the single star sequence ($m_{\mathrm{s}}$) and one for the binary star sequence ($m_{\mathrm{b}}$).

\item One parameter $a$ setting the (linear) slope of the field star brightness distribution within the bin, dN/d($\Delta m$). By defining the field star distribution in terms of height above/below the cluster sequence rather than in terms of $i$, we linearize what is otherwise a two-dimensional function of $i$ and $(g-i)$.
\end{enumerate}

The PDFs for the cluster single star population, cluster binary star population, and field population are therefore given, in units of stars per unit area per unit magnitude, by

\begin{equation}\label{eq:phi_s}
 \phi_{\mathrm{s}} (m,r) = N_{s} \times [\frac{1}{2 \pi r_0^2} \exp(-\frac{r}{r_0})] \times [\frac{1}{\sigma_{\mathrm{s}} \sqrt{2\pi}}\exp(-\frac{(m-m_{\mathrm{s}})^2}{2\sigma_{\mathrm{s}}^2})],
\end{equation}

\begin{equation}\label{eq:phi_b}
 \phi_{\mathrm{b}} (m,r) = N_{b} \times [\frac{1}{2 \pi r_0^2} \exp(-\frac{r}{r_0})] \times [\frac{1}{\sigma_{\mathrm{b}} \sqrt{2\pi}}\exp(-\frac{(m-m_{\mathrm{b}})^2}{2\sigma_{\mathrm{b}}^2})], \mathrm{and}
\end{equation}

\begin{equation}\label{eq:phi_f}
 \phi_{\mathrm{f}} (m,r) = [N_{\mathrm{tot}} - N_{s} - N_{b}] \times [\frac{1}{A_{\mathrm{tot}}}] \times [a (m-\frac{1}{4}) + \frac{1}{2}].
\end{equation}
Note that the final term in Equation~\ref{eq:phi_f}, which treats the magnitude dependence linearly, contains constants of normalization that change depending on the interval over which the field population is being fitted. This calculation assumes that the lower limit is $-0.75$ mag and the upper limit is $+1.25$ mag.

\subsection{Application of the Method to M37}\label{mem_M37}
We considered all stars in a color bin that fell within 0.75 mag below or 1.25 mag above the MS. We fit for the values for all of these parameters in each color bin using a random walk around an initial set of guessed parameter values to find a minimum in $\chi^2$ space. We used

0.2 < $(g-i) \leq$ 0.7   (F5-G8),

0.7 < $(g-i) \leq$ 1.4   (G8-K5),

1.4 < $(g-i) \leq$ 2.1   (K5-M0),

2.1 < $(g-i) \leq$ 2.8   (M0-M2.5),

2.8 < $(g-i) \leq$ 3.2   (M2.5-M4.5), and

\begin{figure}
\centerline{\includegraphics{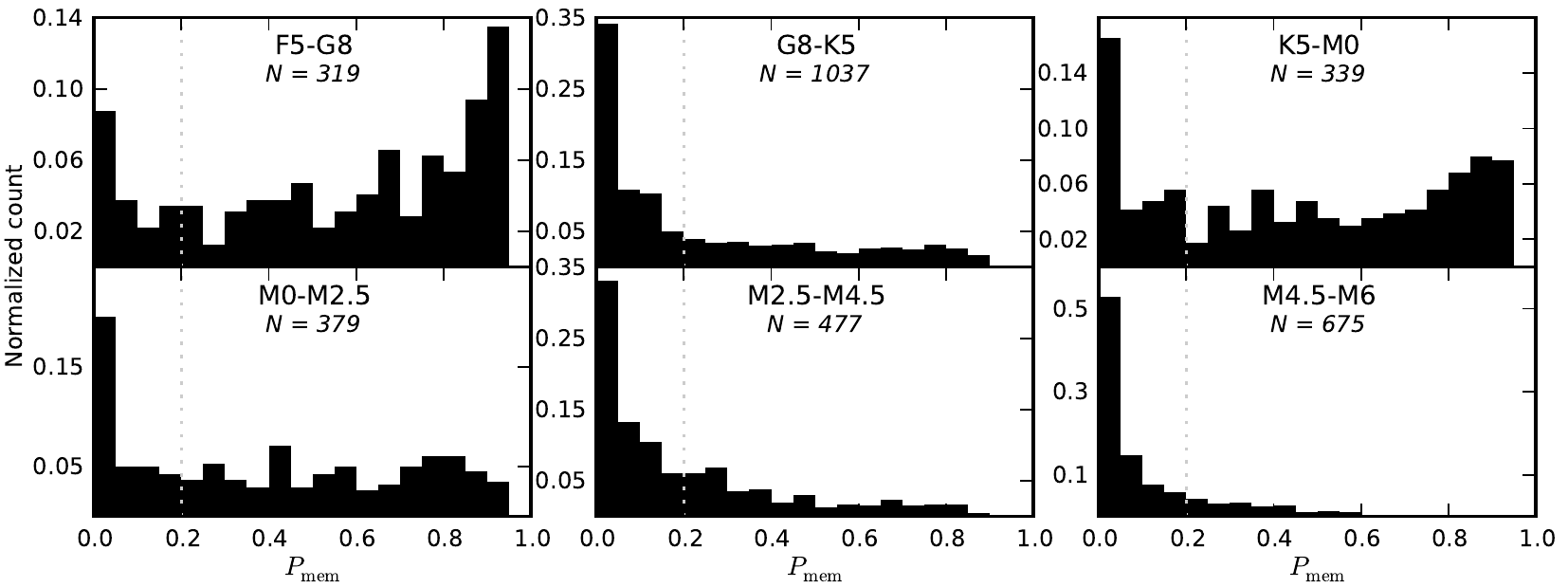}}
\caption{Normalized histograms of the number of cluster candidates ($P_{\mathrm{mem}}>0$) as a function of $P_{\mathrm{mem}}$ for each color bin. The total number of stars $N$ in each color bin is shown. The vertical dotted line indicates the value of $P_{\mathrm{mem}}=0.2$ that we use in our analysis of M37 to separate cluster member from non-members.}
\label{fig:pmems_hist}
\vspace{0.1 in}
\end{figure}

3.2 < $(g-i) \leq$ 3.6   (M4.5-M6)

\noindent as the color bins covering the range of the discernible MS. We then calculated three probabilities to each star: one of being an isolated member ($P_{\mathrm{s}}$), one of  being likely a binary ($P_{\mathrm{b}}$), and one of being a field star ($P_{\mathrm{f}}$). Note that $P_{\mathrm{s}} + P_{\mathrm{b}} + P_{\mathrm{f}} = 1.0$ for every star. Table~\ref{tbl:cluster_fit} summarizes the parameters of Equations~\ref{eq:phi_s}, \ref{eq:phi_b}, and \ref{eq:phi_f} obtained with the fit, and Figure~\ref{fig:pmems_hist} shows a histogram of the number of candidates as a function of $P_{\mathrm{mem}}$ for each color bin.

\begin{deluxetable}{@{}cc*{10}{r}ccc@{}}
\tabletypesize{\scriptsize} 

\tablecaption{Cluster Fit Parameters for M37 \label{tbl:cluster_fit}}

\tablehead{
\colhead{SpType}           & \colhead{$N_{\mathrm{f}}$} & \colhead{$N_{\mathrm{s}}$} &
\colhead{$N_{\mathrm{b}}$} & \colhead{$r_0$}            & \colhead{$a$}              &
\colhead{$m_{\mathrm{s}}$} & \colhead{$\sigma_{\mathrm{s}}$} & \colhead{$m_{\mathrm{b}}$} &
\colhead{$\sigma_{\mathrm{b}}$} \\
 & & & & (deg) & & (mag) & & (mag) & \\[-0.1 in]
}
\startdata
F5.0-G8.0 & 137 & 165 & 23 & 0.044 & -0.238 & 0.010 & 0.181 & 0.636 & 0.260 \\
G8.0-K5.0 & 915 & 162 & 87 & 0.043 & -0.395 & 0.018 & 0.103 & 0.373 & 0.306 \\
K5.0-M0.0 & 446 & 156 &  0 & 0.054 & -0.309 & 0.027 & 0.068 & 0.357 & 0.347 \\
M0.0-M2.5 & 468 & 153 &  0 & 0.053 & -0.348 & 0.017 & 0.084 & 0.535 & 0.448 \\
M2.5-M4.5 & 647 &  96 &  0 & 0.050 & -0.394 & 0.050 & 0.119 & 0.455 & 0.058 \\
M4.5-M6.0 & 940 &  18 & 53 & 0.046 & -0.455 & 0.017 & 0.064 & 0.244 & 0.192 \\[-0.1 in]
\enddata
\vspace{0.1 in}
\end{deluxetable}

\begin{figure}
\centerline{\includegraphics{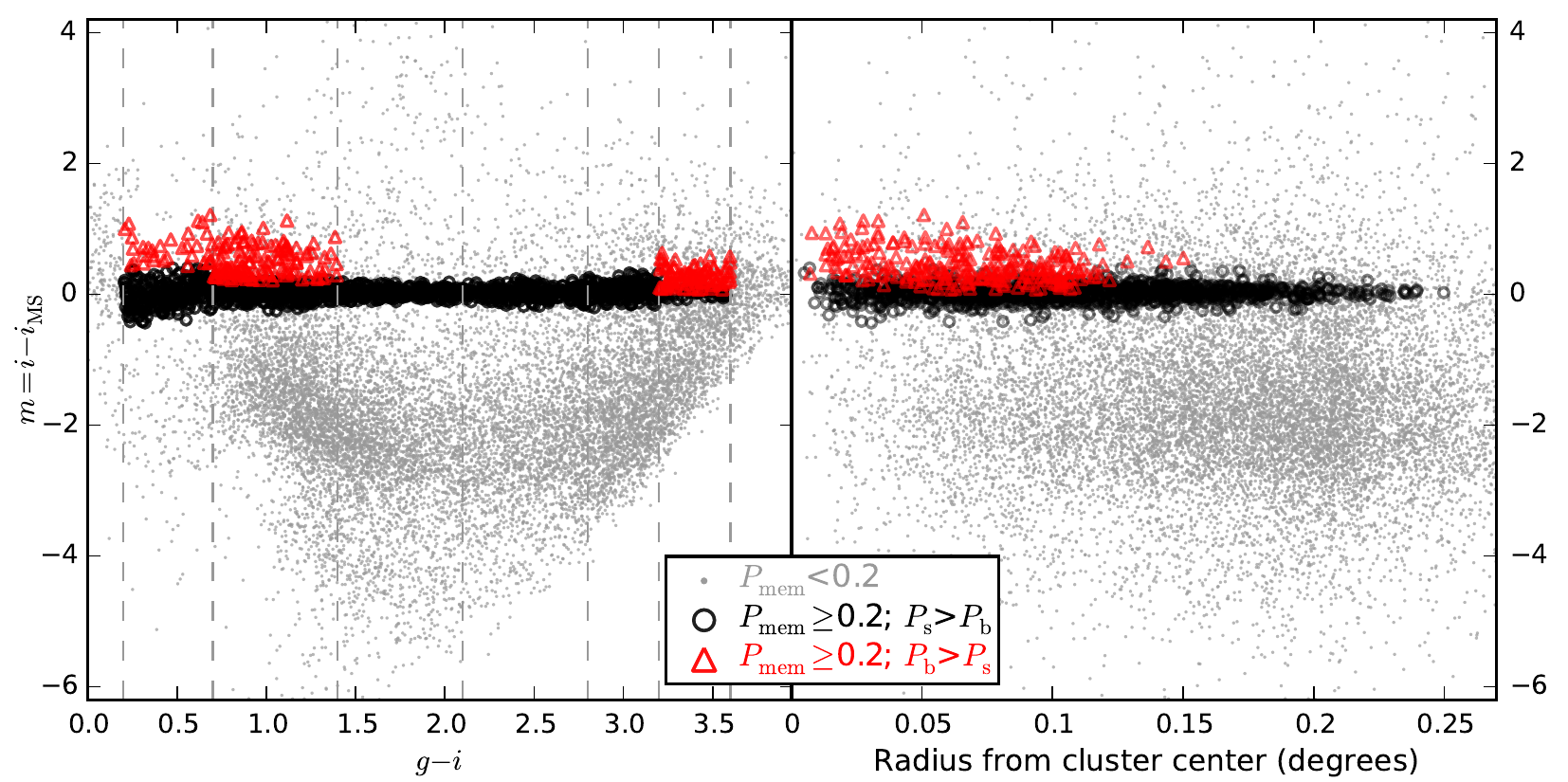}}
\caption{$Left$ --- Distance $m$ above/below the MS of M37 in the ($i$,$(g-i)$) CMD as a function of $(g-i)$ color for stars in the M37 field. We color-code stars using our calculated \Pmem: field stars (\Pmem\ < 0.2) in gray, single members (\Pmem\ $\geq 0.2$ and $P_{\mathrm{s}}>P_{\mathrm{b}}$) in black, and likely binary members (\Pmem\ $\geq 0.2$ and $P_{\mathrm{b}}>P_{\mathrm{s}}$) in red. Vertical dashed lines indicate the limits of the color bins used in the analysis. $Right$ --- Same as left panel, with distance from the MS of M37 as a function of radius from cluster center. Binaries appear to be concentrated toward the center of the cluster.}
\label{fig:memMS}
\end{figure}

We observe in the left panel of Figure \ref{fig:memMS} a segregation of likely binaries toward the blue and red ends of the MS. The fact that most stars in the red end of the sequence are considered binaries illustrates a possible sign of mass segregation and tidal evaporation of the low-mass single stars, so that only the more massive binaries are left. This mass segregation and preferential loss of low-mass single members has been seen in many other clusters, including Praesepe \citep{pinfield2003} and the Pleiades \citep{lodieu2007} where the binary fraction increases in the lowest-mass bins in a way not seen for cool field stars \citep[e.g.,][]{duchene2013}. The lack of binaries at 1.4 < $(g-i)$ < 3.2 (K5--M4.5) suggests that the binary sequence is not sufficiently well detected to be fit as a distinct component by our algorithm. The small number of over-luminous binary systems is likely being incorporated into the fit for the field population, indicating one possible source of incompleteness.

Lastly, we treated stars blue of the turn-off point (at $(g-i)\approx0.2$) differently because the MS there is almost vertical in ($g-i$) and, thus, our standard approach cannot be applied. Instead of having height $m$ above or below the MS as one of the membership criteria, we considered as a single group all stars within the following CMD locus (with no further consideration of CMD position or single/binary flag): $-0.3<(g-i)<0.2$ and brighter than 0.75 mag below the MS, and $0.2<(g-i)<0.5$ and brighter than 1.25 mag above the MS. This locus catches stars blue of the turn-off and those moving across the Hertzsprung gap. We then fitted using the spatial term as before to factor into the membership probability. Applying this approach, we found that of the 198 stars in the locus, 171 have \Pmem\ $\geq$ 0.2, 139 have \Pmem\ $\geq$ 0.5, and 74 have \Pmem\ $\geq$ 0.8. The scale radius is 0$^{\circ}$.038, which is smaller than that of the redder bins (see Table~\ref{tbl:cluster_fit}). This is consistent with the suggestion of mass segregation seen in the rest of the cluster members.

\subsection{Testing the Method with Praesepe}\label{mem_test}
To test the robustness of the membership probability method described above, we implemented it on stars in and around the Praesepe open cluster. \citet{adam2007} calculated Praesepe cluster \Pmem\ using both photometric and proper motion data to identify $\approx$1000 members from among the several million stars they examined.

To implement our membership calculation, which excludes proper motion information, we limited our sample of Praesepe point sources to a circular locus in the proper motion diagram centered around the mean cluster proper motion. This was needed to obtain a sample set with a members-to-contaminants ratio similar to that of the M37 field. We also limited our sample to objects in the range $2.8<(g-i)<3.2$ (spectral types $\approx$M2.6--M4.3,) which gives us enough cluster members to compare.

\begin{figure}
\centerline{\includegraphics{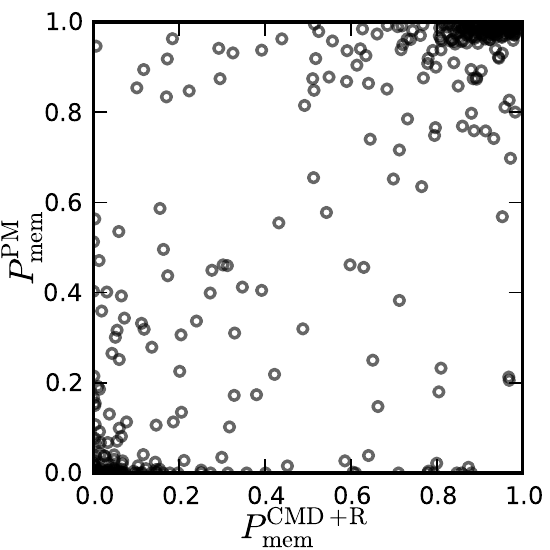}}
\caption{Comparison between two different sets of Praesepe cluster \Pmem\ calculations: one using CMD distance, radial distance, and proper motion information ($P_{\mathrm{mem}}^{\mathrm{PM}}$), as calculated by \citet{adam2007}, and another one using only CMD and radial distances ($P_{\mathrm{mem}}^{\mathrm{CMD+R}}$), as calculated here. As expected, the overwhelming majority of stars have comparable \Pmem: likely members are in the upper right locus, and non-members in the lower left locus. The subset of stars with high $P_{\mathrm{mem}}^{\mathrm{PM}}$ and low $P_{\mathrm{mem}}^{\mathrm{CMD+R}}$ (upper left) are likely part of the wings of the actual member distribution.}
\label{fig:mem_test}
\vspace{0.1 in}
\end{figure}

The results of our own CMD+radial-distance-based membership probability calculations ($P_{\mathrm{mem}}^{\mathrm{CMD+R}}$) are shown in Figure~\ref{fig:mem_test}, where we compare them against proper-motion-based calculations by \citet{adam2007} ($P_{\mathrm{mem}}^{\mathrm{PM}}$) for the same stars. We found 296 objects with $P_{\mathrm{mem}}^{\mathrm{CMD+R}}$ > 0.6 (232 single stars and 64 binaries); these are high-confidence Praesepe members. \citet{adam2007} found 310 members in the same photometric range.

The expectation that stars with high $P_{\mathrm{mem}}^{\mathrm{PM}}$ should also have high $P_{\mathrm{mem}}^{\mathrm{CMD+R}}$ appears to hold in practice. Likely members flagged as such by both calculations are in the upper right of Figure~\ref{fig:mem_test}, while likely field stars are in the lower left. Sources in the upper left (high $P_{\mathrm{mem}}^{\mathrm{PM}}$, low $P_{\mathrm{mem}}^{\mathrm{CMD+R}}$) may be either cluster members on the wings of the photometric distribution or interlopers slightly in front or behind the cluster. The progressive way in which this subset of members smoothly meets the member locus in the upper right convinced us that these are most likely part of the wings of the actual member distribution. Objects in the lower right (low $P_{\mathrm{mem}}^{\mathrm{PM}}$, high $P_{\mathrm{mem}}^{\mathrm{CMD+R}}$) are probably non-member dwarfs in the same spatial volume considered in the analysis. In summary, we deem our membership probability calculation to be robust.

\section{Spectral and Timing analysis for some high-count X-ray sources}\label{spectra}
We performed spectral and timing analysis on the four highest-count sources in our \CHANDRA\ catalog: XID 65 (1267 total source counts in the full band), XID 108 (3097), XID 144 (4398), and XID 1004 (5814). Only XID 65 and XID 108 have counterparts on our catalog of optical objects (see Section~\ref{consolidation},) but neither one is a likely cluster member. Source XID 108 is matched to the eclipsing binary of W UMa type KV11, which \citet{kang07} characterized as having a period of 1.89 days and a $V$ magnitude amplitude of 0.15 mag. On the other hand, the only knowledge we have of the optical counterpart to source XID 65 is its $gri$ photometry. Table~\ref{tbl:X_opt} shows the characteristics of the optical counterparts to the high-count X-ray sources, as well as the hardness ratio of the matched X-ray source.

\begin{figure}
\centerline{\includegraphics{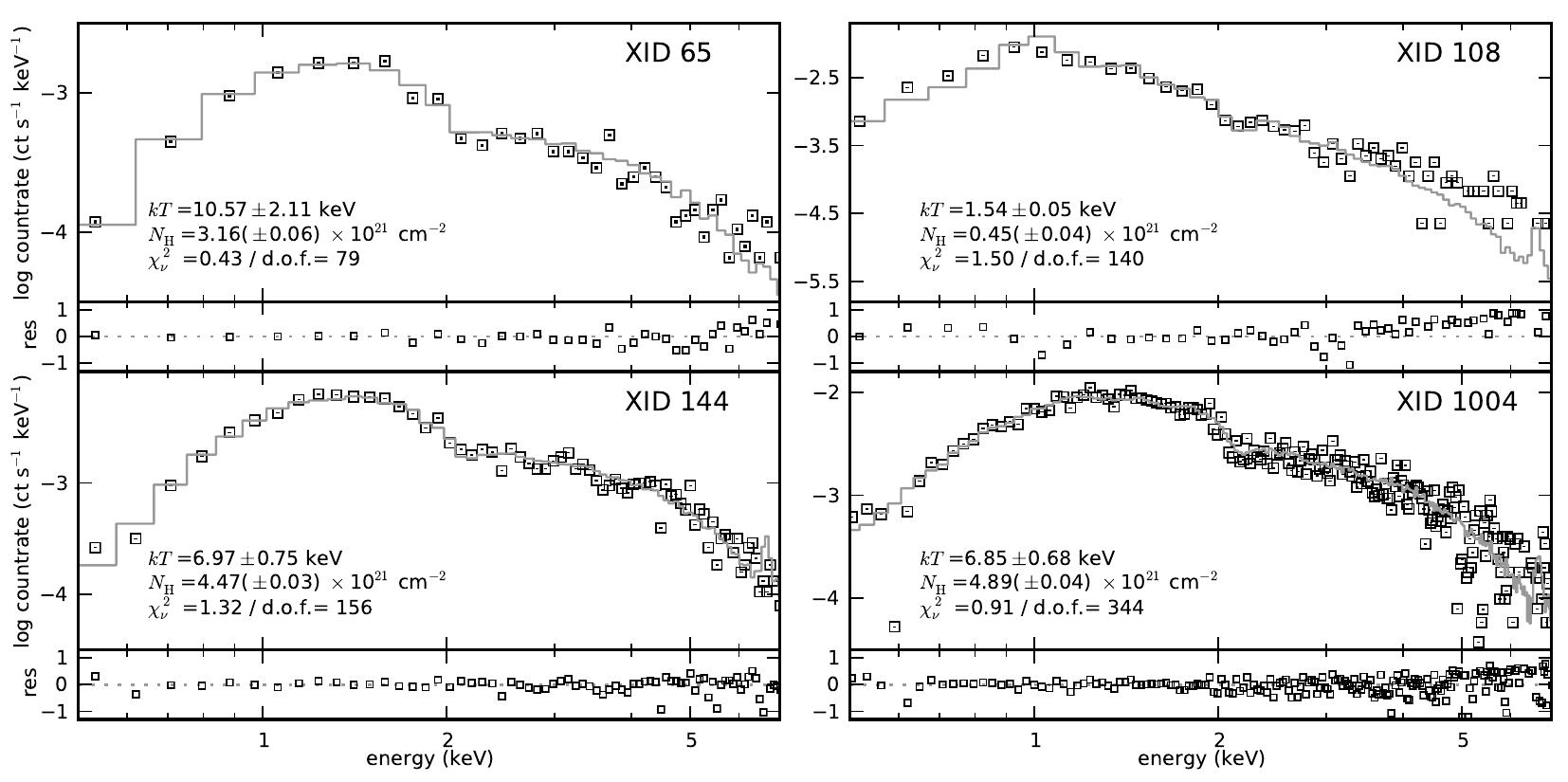}}
\caption{Spectral fits for the brightest sources in our \CHANDRA\ catalog. The 1T-APEC model fit for each spectrum (solid gray line) is drawn over the binned (15--25 counts per bin) data points. The error bars of the binned data points are too small to show. The parameter values for the fit using the 1T-APEC model are included in each panel: temperature of 1T-APEC model ($kT$), atomic neutral hydrogen column density ($N_{\mathrm{H}}$), reduced chi-squared statistic ($\chi^2_{\nu}$), and degrees of freedom (d.o.f.). The residuals of the fit are shown at the bottom of each panel, normalized to the source counts. The source spectrum and spectral fit of XID 108 exclude events during a multi-standard deviation flare, as seen in Figure~\ref{fig:lightcurves}.}
\label{fig:specs}
\end{figure}

We used the CIAO tool $Sherpa$\footnote{\url{http://cxc.harvard.edu/sherpa/}} to fit each spectra with three different $xspec$ models using solar abundances by \citet{wilms2000}): one-temperature (1T) and two-temperature (2T) APEC models, and a power law model. APEC models are used to model the X-ray emission of diffuse gas in collisional equilibrium, characteristic of stellar coronae. The power-law model is a simple photon power law of the form $A(E) = KE^{-\Gamma}$, where $K$ is a normalization constant. We used the ISM absorption model $tbabs$ using photoelectric cross-section from \citet{balucinska1992} to account for extinction by atomic neutral hydrogen. Since the distance to these X-ray sources is undetermined, we left the atomic neutral hydrogen column density ($N_{\mathrm{H}}$) as a free parameter. For each source, we fitted the spectra of the five individual observations simultaneously using the chi-squared ($\chi^2$) statistic with the Gehrels variance function and $Sherpa$'s $levmar$ optimization method. We obtained 1$\sigma$ or 68\% confidence intervals of all free parameters by computing the covariance matrices. In Figure~\ref{fig:specs} we compare the binned spectra (15 to 25 counts per bin) of the sources against a best fit 1T-APEC model. We report in Table~\ref{tbl:spectra} all acceptable fits for each X-ray source.

\begin{deluxetable*}{rrrrrcccc}
\tabletypesize{\scriptsize}

\tablecaption{Properties of Optical Counterparts to High-Count X-ray Sources
\label{tbl:X_opt}}

\tablehead{
\colhead{XID} & \colhead{Opt. ID} & \colhead{$g$} &
\colhead{$r$} & \colhead{$i$}     & \colhead{$V$} &
\colhead{$B-V$} & \colhead{$F_{\mathrm{X}}/F_{\mathrm{opt}}$\tablenotemark{a}} &
\colhead{$HR$} \\
\colhead{}      & \colhead{}      & \colhead{(mag)} &
\colhead{(mag)} & \colhead{(mag)} & \colhead{(mag)} & \colhead{(mag)} & \\[-0.12 in]
\setcounter{column_number}{1}
}

\startdata
65   & 240271 & 21.53 & 20.76 & 20.46 & 21.06\tablenotemark{b} & \nodata & 2.88 & -0.12\\
108  & 220858 & 15.08 & 14.15 & 13.72 & 14.56   & 0.66    & 0.01 & -0.72 \\[-0.1 in]
\enddata
\tablenotetext{a}{Ratio derived using the absorbed X-ray flux in the 0.5--7 keV band and the flux ratio definition by \citet{stocke91}.}
\tablenotetext{b}{Derived using the $gr$-to-$V$ transformation by \citet{jester2005}.}
\end{deluxetable*}

\begin{deluxetable*}{lcrrrrrrl ccc@{}}
\tabletypesize{\scriptsize}

\tablecaption{Spectral Fits for the Highest-Count X-ray Sources in our {\em Chandra} Catalog\label{tbl:spectra}}

\tablehead{
\colhead{XID}    & \colhead{Model}           & \colhead{$kT_1$}           &
\colhead{$kT_2$} & \colhead{$\Gamma$}        & \colhead{$N_{\mathrm{H}}$} &
\colhead{Flux}   & \colhead{$\chi^2_{\nu}$/d.o.f.} & \colhead{Note}\\
\colhead{}       & \colhead{}                & \colhead{(keV)}            &
\colhead{(keV)}  & \colhead{}                & \colhead{(cm$^{-2}$)}      &
\colhead{(erg cm$^{-2}$ s$^{-1}$)}           & \colhead{}                 & \colhead{} \\
 & & & & & \multicolumn{1}{c}{$\times 10^{21}$} &
\multicolumn{1}{c}{$\times 10^{-14}$}           & & \\[-0.12 in]
\setcounter{column_number}{1}
}

\startdata
65   & 1T & 10.57$\pm$2.11 &\nodata &\nodata & 3.16$\pm$0.60 & 5.44$\pm$0.27  & 0.43/79 & \nodata \\
     & PL & \nodata        &\nodata & 1.56$\pm$0.12 & 3.77$\pm$1.01 & 5.59$\pm$0.32  & 0.44/79 & \nodata \\
108  & 1T & 1.54$\pm$0.05  &\nodata &\nodata & 0.45$\pm$0.45 & 5.19$\pm$0.28  & 1.50/140 & Flare subtracted\\
     & PL & \nodata        &\nodata & 3.59$\pm$0.17 & 3.76$\pm$0.48 & 13.31$\pm$1.72 & 1.19/140 & Flare subtracted\\
     & 2T & 0.79$\pm$0.10  & 3.77$\pm$1.91 &\nodata & 6.25$\pm$1.13 & 41.00$\pm$5.69 & 0.69/25  & Flare only\\ 
     & PL & \nodata        &\nodata & 2.60$\pm$0.23 & 3.04$\pm$1.00 & 30.90$\pm$4.28 & 0.69/26  & Flare only\\
144  & 1T & 6.97$\pm$0.75  &\nodata &\nodata & 4.47$\pm$0.33 & 19.03$\pm$0.49 & 1.32/156 & \nodata \\
     & PL & \nodata        &\nodata & 1.79$\pm$0.07 & 5.83$\pm$0.56 & 20.51$\pm$0.72 & 1.31/156 & \nodata \\
1004 & 1T & 6.84$\pm$0.68  &\nodata &\nodata & 4.89$\pm$0.35 & 27.18$\pm$0.73 & 0.91/344 & Background subtracted\\
     & 2T & 6.26$\pm$0.67  & 0.34$\pm$0.18 &\nodata & 5.94$\pm$0.74 & 29.29$\pm$1.12 & 0.91/343 & Background subtracted\\
     & PL & \nodata        &\nodata & 1.78$\pm$0.07 & 6.10$\pm$0.48 & 29.92$\pm$0.90 & 0.91/344 & Background subtracted\\[-0.1 in]
\enddata
\tablecomments{Results of all acceptable spectral fits for our four highest-count X-ray sources. For the one-temperature (1T) and two-temperature (2T) APEC models we give the temperature components $kT_1$ and $kT_2$. For the power law (PL) model we give the photon index $\Gamma$. All fits included the $xspec$ model $tbabs$ to account for ISM extinction, and we give the best fit atomic hydrogen column density $N_{\mathrm{H}}$. We also show the derived unabsorbed flux in the 0.5--7 keV band. Finally, we give for each fit its reduced chi-square ($\chi^2_{\nu}$) and degrees of freedom (d.o.f.).
}
\end{deluxetable*}
\vspace{0.2 in}

We only subtracted background counts on source XID 1004, as this source is located on the back-illuminated, more background sensitive ACIS-S3 chip. The ratio between background and net count rates for XID 1004 is $\approx$40\%. For the other three sources, their location on the front-illuminated ACIS-I chips guarantees that background noise remains low. The ratio between background and net count rates is 8.5\% for XID 65, 3.6\% for XID 108, and 4.2\% for XID 144.

\begin{figure}
\centerline{\includegraphics{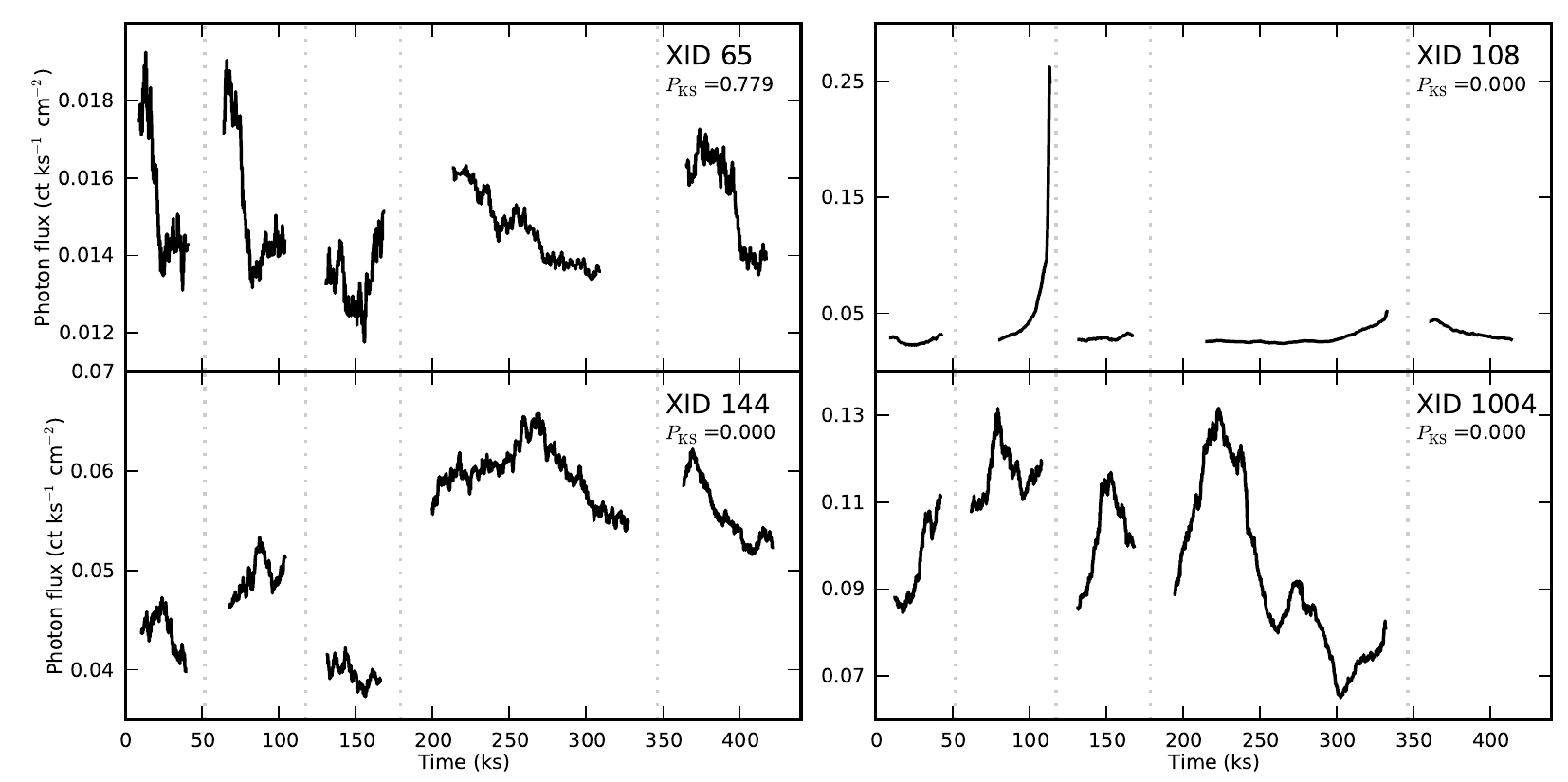}}
\caption{Adaptively smoothed light curves of the brightest sources in our \CHANDRA\ catalog in the 0.5--7 keV energy range. Background counts are not subtracted. Vertical dotted lines represent the cutoffs between our different \CHANDRA\ observations. The light curves from the five observations are in chronological order from left to right, and the time between observations is ignored. Note that source XID 1004 was not in the last observation (ObsID 13657). The $P_{\mathrm{KS}}$ statistic of variability is shown for each source. $P_{\mathrm{KS}}>0.05$ indicates no variability; $P_{\mathrm{KS}}<0.005$ indicates definite variability. Source XID 108 displays a multi-standard deviation flare. We extracted two spectra from the light curve of this source: one excluding events during the flare, and one including only flare events.}
\label{fig:lightcurves}
\end{figure}

We used AE's tools to extract adaptively smoothed light curves of the four sources. These tools also perform a test of time variability by comparing a uniform flux model to the distribution of source event time stamps and then computing a 1-sample Kolmogorov-Smirnov statistic $P_{\mathrm{KS}}$. A statistic value of $P_{\mathrm{KS}}>0.05$ means there is no evidence for variability, while $P_{\mathrm{KS}}<0.005$ indicates that the source is definitely variable. Only XID 65 had $P_{\mathrm{KS}}>0.05$, suggesting that it is the only one of the four sources with no evidence of variability. In Figure~\ref{fig:lightcurves} we put in sequence the light curves of our five \CHANDRA\ observations for each source, and we show $P_{\mathrm{KS}}$ for each source. The vertical dotted lines indicate the cutoffs between observations, and the time between observations is not included. Note that source XID 1004 was not in the last observation (ObsID 13657). 

We used AE's \texttt{ae\_pileup\_screening} tool to check for potential photon pile-up in these four high-count sources. The procedure calculates on the individual observations the number of counts per ACIS frame using 3x3 pixel islands within the PSF of the source. We found that the highest 3x3 rate was 0.02 counts per frame---corresponding to a $\approx$1\% pile-up fraction---for XID 108, and 0.01 counts per frame for XID 144. These pile-up fractions are below the suggested 0.05 counts-per-frame value above which pile-up reconstruction is warranted. Sources XID 65 and 1004 had a 3x3 rate of zero.

There is a multi-standard deviation flare in source XID 108, as evident on its light curve in Figure~\ref{fig:lightcurves}. Including this flare in the spectra of the source led to unacceptable spectral fits ($\chi^2_{\nu}>2$) on all the models we tried. We thus fitted the spectrum of the flare separately from the rest of the spectrum of XID 108, and we included both fits in Table~\ref{tbl:spectra}. 

All four spectra are well-fitted using either a 1T-APEC model or a power-law model. Only source XID 1004 and the flare of XID 108 had a 2T-APEC model fit that resulted in two $kT$ temperatures significantly above zero. Finally, the $N_{\mathrm{H}}$ values derived by the $tbabs$ model for the four X-ray sources are larger than the 1.26$\times 10^{21}$ cm$^{-2}$ for M37 (derived from optical reddening, see Section~\ref{quantities-xray}), suggesting that these four high-count sources may be background sources. The only exception is one of the fits for XID 108: $N_{\mathrm{H}}$ from the 1T-APEC model fit suggests that it is a foreground object, whereas $N_{\mathrm{H}}$ from the power-law model fit suggests that it is a background object. The web-based FTOOL $nh$\footnote{\url{http://heasarc.gsfc.nasa.gov/cgi-bin/Tools/w3nh/w3nh.pl}} indicates a total Galactic $N_{\mathrm{H}}$ of $\approx$4$\times 10^{21}$ cm$^{-2}$ in the direction of M37 based on the Leiden/Argentine/Bonn survey of Galactic atomic hydrogen. This means that sources XID 144 and XID 1004, having $N_{\mathrm{H}}$ larger than that, are most likely extragalactic sources.

\end{document}